# Simultaneous Sequential Detection of Multiple Interacting Faults

Ram Rajagopal[1], XuanLong Nguyen[2], Sinem Coleri Ergen[3] and Pravin Varaiya[1]

[1] Electrical Engineering and Computer Sciences, University of California, Berkeley, USA
[2] Statistics, University of Michigan, Ann Arbor, USA
[3] Koc University, Istanbul, Turkey
{ramr,varaiya}@eecs.berkeley.edu, xuanlong@umich.edu, sergen@ku.edu.tr

**Abstract**

Single fault sequential change point problems have become important in modeling for various phenomena in large distributed systems, such as sensor networks. But such systems in many situations present multiple interacting faults. For example, individual sensors in a network may fail and detection is performed by comparing measurements between sensors, resulting in statistical dependency among faults. We present a new formulation for multiple interacting faults in a distributed system. The formulation includes specifications of how individual subsystems composing the large system may fail, the information that can be shared among these subsystems and the interaction pattern between faults. We then specify a new sequential algorithm for detecting these faults. The main feature of the algorithm is that it uses composite stopping rules for a subsystem that depend on the decision of other subsystems. We provide asymptotic false alarm and detection delay analysis for this algorithm in the Bayesian setting and show that under certain conditions the algorithm is optimal. The analysis methodology relies on novel detailed comparison techniques between stopping times. We validate the approach with some simulations.

## I. INTRODUCTION

Sequential change point detection problems have been widely studied [12] when involving a single fault or multiple hypothesis based on a single change. New large distributed systems exhibit fault behaviors that required modeling of multiple correlated faults [4]. For example, in a sensor network each sensor can fail independently of each other, and the correlation between pairs of sensors can be used for diagnosis (e.g. see [21]). The faults are interacting since a fault in any pair of sensors causes a change in the correlation between them. In this paper we are concerned with the problem of detecting multiple interacting faults. This requires a new formulation that differs from the single fault problem.

**Single faults**. Classic sequential change point detection [12] is concerned with variations on the following basic problem: given a sequence of random observations $\{X_k, k \geq 1\}$, such that $X_k$ is distributed with density $f_0$ (i.e. $X_k \sim f_0$) if $k < \lambda$ and $X_k \sim f_1$ if $k \geq \lambda$ for a random change time $\lambda \sim \pi$, find a procedure $\bar{\nu}$ that detects and stops at time $n$ if $\lambda \leq n$ on the basis of the observations $\mathcal{F}_n(X) = \{X_k, 1 \leq k \leq n\}$. The change behavior can be compactly denoted by $f_0 \xrightarrow{\lambda} f_1$. Various solutions have been proposed for this problem, such as the CUSUM [18] and the Shiryaev-Roberts-Pollak (SRP) [22], [23] procedures.

Research supported by California Department of Transportation and ARO-MURI UCSC-W911NF-05-1-0246-VA-09/05.

The gist of these approaches is a threshold test for the likelihood ratio at time $n$

$$\Lambda_n(X) = \frac{\mathbb{P}(\lambda \leq n \,|\, \mathcal{F}_n(X))}{\mathbb{P}(\lambda > n \,|\, \mathcal{F}_n(X))} \geq B_\alpha, \tag{1}$$

when using the complete Bayesian model. The threshold $B_\alpha$ is chosen to satisfy a false alarm constraint $\mathbb{P}(\bar{\nu} < \lambda) \leq \alpha$. The procedure can be defined as the stopping time $\bar{\nu}$ such that

$$\bar{\nu} = \inf\{n : \Lambda_n(X) \geq B_\alpha\}, \tag{2}$$

and its performance is measured by the m-moment of detection delay

$$D_m^\lambda(\bar{\nu}) = \mathbb{E}_\lambda\left[(\bar{\nu} - \lambda)^m \,|\, \bar{\nu} \geq \lambda\right],$$

where $\mathbb{E}_\lambda$ denotes expectation with respect to the prior of $\lambda$ and typically $m = 1$ or $m = 2$. Asymptotic performance of single change point procedures in this and other performance criteria have been extensively analyzed (e.g. [2], [11], [13], [20], [25]). In particular, Tartakovsky et al. [25] show the asymptotic delay optimality of the SRP rule under diminishing false alarm probability and threshold

$$B_\alpha = \frac{1-\alpha}{\alpha} \tag{3}$$

is

$$D_m^\lambda(\bar{\nu}) \doteq \left[\frac{|\log \alpha|}{q_1(X) + d}\right]^m, \tag{4}$$

where $\doteq$ denotes asymptotic upper and lower bounds with respect to $\alpha \to 0$. The delay is only a function of the false alarm $\mathbb{P}(\lambda < \bar{\nu}) \leq \alpha$, the amount of information $q(X)$ in the densities $f_0$ and $f_1$ and the tail exponent $d$ of the prior for $\lambda$:

$$q_1(X) = \int f_1(x) \log \frac{f_1(x)}{f_0(x)} \mu(dx).$$

$D_m^\lambda(\bar{\nu})$ is also the minimum asymptotic delay achievable by *any* procedure with false alarm $\alpha$. The single change point model captures problems of fault diagnosis, where the measured data is fully observed and the change in the measurements is attribute to a single fault happening at a random time.

**Multiple simultaneous interacting faults** can happen in a complex system with multiple interacting components. Consider the system in Figure 1(a). Each node in Figure 1(a) is a subsystem and each edge represents information shared between subsystems. There are multiple subsystems, $u_1$ to $u_5$, each of which can fail at random times $\lambda_1$ to $\lambda_5$. A sequence of observations $X_n(u_i)$ is collected at each subsystem $i$. When subsystem $u_i$ fails, the sequence $X_n(u_i)$ experiences a change. Since this sequence is only collected by an individual subsystem its denominated *private information* of subsystem $u_i$. Moreover subsystems $u_i$ and $u_j$ also collect a shared sequences of observations $X_n(u_i, u_j)$ that is influenced by failures in either subsystem. These sequences are denominated *shared information* between subsystems $u_i$ and $u_j$. Since the graph in Figure 1(a) specifies the pattern of information sharing among subsystems we denominate it *communication graph*. Information could be shared by more than two subsystems, and would be represented by a hyper-edge connecting multiple nodes in the graph.

Solving the multiple interacting fault detection problem requires creating a test for each subsystem to detect its own failure using only the local information it collects, namely the private and shared information available to it. Each subsystem could use its private information sequence $X_n(u_i)$ and the SRP rule in Eq. (2) to obtain a stopping time to detect its own fault. But clearly this procedure does not use and benefit from the shared information at all.

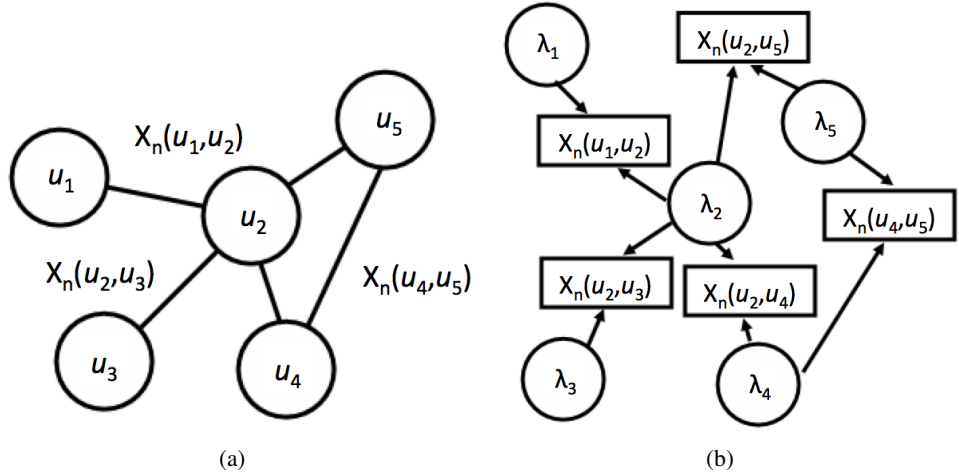

Fig. 1. (a) Communication graph: information sharing graph of a system with multiple subsystems (nodes) with edges between nodes indicating shared information and (b) Fault graph: statistical dependency between information and failure times.

The shared information can only be used if some structure on how faults interact with shared information is specified. The fault graph (Figure 1(b)) displays graphically the statistical dependency between shared information variables $X_n(u_i, u_j)$ and faults $\lambda_i$ and $\lambda_j$. It is natural for many practical situations to assume that when either subsystem $u_i$ or $u_j$ fails, the shared observation sequence experiences a change in distribution. Furthermore, after one of the subsystems has failed, the shared information relating to that subsystem becomes useless to detect a fault on the other subsystems using the same shared information. Therefore, the earliest of the fault times $\lambda_i$ and $\lambda_j$ drives a change in the distribution of $X_n(u_i, u_j)$. In general situations, alternative functional behaviors could be specified.

The interaction of faults in shared information makes it very challenging to use this information in a test for a subsystem. For example, a very naive test that only used a single sequence $X_n(u_i, u_j)$ to diagnose subsystem $u_i$ would be driven to an incorrect decision if subsystem $u_j$ fails long before subsystem $u_i$ fails. Thus the integration of weak evidence to build an effective detection procedure is required.

One useful practical application of the stochastic model we discuss is in detecting faulty sensors in a sensor network measuring a slowly varying spatial and temporal process. Each individual sensor in a network can fail at some random unknown time. The nature of failure is such that plausible measurements are still reported. Sensors deployed geographically near each other compare their information to determine whether they are failed or not. Before failure, measurements maintain some degree of similarity due to the slow varying nature of the phenomena being measured, and after failure this similarity is significantly reduced. In our current setup, each sensor is a subsystem. The private information are similarity comparisons to a sensor's own past measurements or to some reference working sensor. The shared information are similarity comparisons of a sensor to nearby sensors. Various studies [9] have proposed ad-hoc and empirical approaches to this problem, but to the best of our knowledge, no systematic theory has been presented. Empirical validation and the implementation details of a solution for this problem in the context of applications can be found in [3], [21], [29].

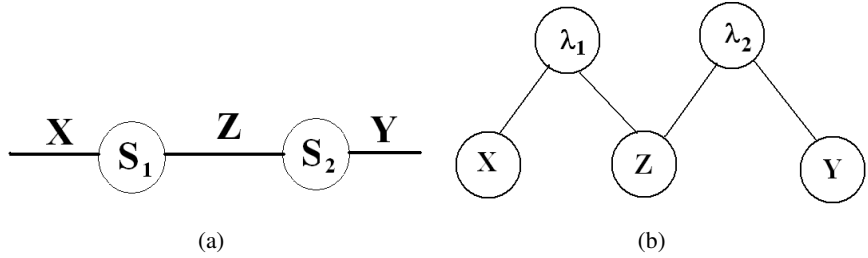

Fig. 2. Interacting subsystems setup: (a) communication graph and (b) fault graph. Information set at time $n$ are denoted $X_n$, $Y_n$ and $Z_n$.

## A. Our contributions

Consider the setup in Figure 2. Two subsystems, $u_1 = 1$ and $u_2 = 2$ fail at times $\lambda_1$ and $\lambda_2$ respectively. The subsystems observe variables reflective of their state according to the communication graph in in Figure 2(a). Subsystem 1 observes a private sequence $X$ that changes its distribution according to $\lambda_1$ and subsystem 2 observes a private sequence $Y$ that changes according to $\lambda_2$. Both subsystems observe the shared information sequence $Z$, whose behavior changes according to the earliest between both failure times. All the observations are independent conditional on the change times. In this paper we will explore the construction of fault detection rules for each subsystem that can effectively use private and shared information. The single fault detection problem in this scenario corresponds to subsystem 1 using only its private information to detect failure $\lambda_1$. The interacting fault problem involving two subsystems presents substantial analytic challenge due to the information constraints and the nature of shared information.

The first natural solution to the problem consists of each subsystem using only its private information to make a decision about its state. In this case, the single fault SRP procedure (Eq. 2) can be used to obtain a stopping rule $\tilde{\nu}_1$ for subsystem 1 with asymptotic delay $D_m^{\lambda_1,\lambda_2}(\tilde{\nu}_1)$ given by Eq. (4). The false alarm of the procedure is bounded by $\alpha$ and the delay is independent of the shared information $Z$. If it was known a priori that $Z$ only changed because of subsystem 1, we could include it in the SRP procedure to obtain a stopping time $\nu_1$ using Eq. (2) with the test ratio

$$\Lambda_n(X,Z) = \frac{\mathbb{P}(\lambda_1 \leq n, \lambda_2 = \infty | \mathcal{F}_n(X,Z))}{\mathbb{P}(\lambda_1 > n, \lambda_2 = \infty | \mathcal{F}_n(X,Z))}. \qquad (5)$$

The resulting delay satisfies $D_m^{\lambda_1,\lambda_2}(\nu_1) \leq D_m^{\lambda_1,\lambda_2}(\tilde{\nu}_1)$, since more information can only help. In fact, this is the smallest delay possible for this problem. But clearly the distribution of the shared information depends on both change times and we need to propose a different strategy.

The optimal single fault procedure uses the posterior probability of a change occurring conditional on the available observations. A natural extension of this procedure to the simultaneous fault problem is to use the posterior probability of change for subsystem 1 conditional on both $X$ and $Z$. This probability can be used in the definition of the single fault procedure (Eq. (2)). The false alarm is guaranteed to be less than $\alpha$. But Theorem 1 surprisingly shows this procedure has an asymptotic delay of at least $D_{\lambda_1,\lambda_2}(\tilde{\nu}_1)$, the delay obtained in a optimal procedure that does not use the shared information $Z$. Therefore, it is not trivial to include shared information in a manner that reduces delay.

Instead we propose a procedure based on the following observation: while neither subsystems have failed, the shared information $Z$ is helpful in diagnosing both, and after failure it is only useful in diagnosing the first subsystem to fail. For subsystem 1, we initially test for its failure *assuming* subsystem 2 is not failed (i.e. $\lambda_2 > n$) using both private information $X$ and shared information $Z$. Similarly we test subsystem 2. If subsystem 2 fails, we switch the test in subsystem 1 to a posterior probability test based only on its private information $X$. The proposed procedure is called *stopping time with information exchange* (STIE) and requires exchanging a single bit of information between subsystems so they can communicate their decisions. We denote it $\bar{\nu}_1$ for subsystem 1 and $\bar{\nu}_2$ for subsystem 2.

The first question is regarding the false alarm of STIE. Theorem 2 shows the false alarm for subsystem 1 is bounded by the sum of $\alpha$ and the *error coupling probability* $\xi^\alpha_{\lambda_1,\lambda_2}(\bar{\nu}_1)$. The error coupling probability captures the probability of subsystem 1 being misled to believe its failed due to a truly failed subsystem 2 taking excessively long to declare a failure. If the stopping times for both subsystems are asymptotically decoupled (i.e. the error coupling probability is smaller than $\alpha$), then we can guarantee a false alarm of less than $\alpha$ for $STIE$. Theorem 3 shows this happens when certain natural relationships hold between the amount of private and shared information. The analysis uses large deviation comparisons of stopping times and is of independent interest.

The remaining question is regarding the delay performance of STIE. Theorem 5 shows STIE achieves an improved asymptotic delay performance as $\alpha \to 0$

$$D_m^{\lambda_1,\lambda_2}(\bar{\nu}_1) = [D_m^{\lambda_1,\lambda_2}(\nu_1)(1-\delta_\alpha) + D_m^{\lambda_1,\lambda_2}(\tilde{\nu}_1)\delta_\alpha](1+o(1)),$$

where

$$D_m^{\lambda_1,\lambda_2}(\nu_1) = \left[\frac{|\log \alpha|}{q(X)+q(Z)+d_1}\right]^m,$$

$$D_m^{\lambda_1,\lambda_2}(\tilde{\nu}_1) = \left[\frac{|\log \alpha|}{q(X)+d_1}\right]^m,$$

and $\delta_\alpha$ is a quantity strictly greater than 0 and less than 1. This quantity reflects how much the shared information benefits subsystem 1 as opposed to subsystem 2. Notice $D_m^{\lambda_1,\lambda_2}(\bar{\nu}_1) < D_m^{\lambda_1,\lambda_2}(\tilde{\nu}_1)$. Theorem 4 then shows that under mild conditions this delay matches the best possible performance for any procedure in an appropriately defined set of procedures with joint false alarm $\alpha$. These conditions are the same required for the error coupling probability to be asymptotically small. This surprising result shows that under mild conditions we can *decouple* the change behavior of the shared information, and obtain an asymptotic optimal procedure for multiple simultaneous interacting fault problem with private and shared information. The proposed solution sheds light into how to construct solutions for other interaction structures.

We conclude the paper with various simulation studies that show the validity of the proposed analytic insights. To the best of our knowledge this is the first paper that studies a multiple simultaneous interacting fault problem with information sharing constraints that impose partial observability at each subsystem.

### B. Related work

Various procedures have been proposed for single fault diagnosis with full observation [1]. The asymptotic performance of these procedures have been analyzed in various papers, under different performance criteria and settings [2], [11], [13], [20], [25].

Information constraints arise naturally in the context of sensor networks. In such systems it is desirable for procedures to only use information from geographically close sensors to limit communication costs and improve network lifetimes. Such constraint leads to a distributed processing requirement for single fault problems. Various authors [15], [16], [27], [28] analyze distributed versions of single change point problems, and derive an optimal rule for some cases. To the best of our knowledge, this is the first paper that introduces a model with *multiple interacting* change points and diagnosis restricted by a partial observability condition, both constraints that are important in practice.

In contrast, multiple simultaneous fault problems have been less studied. Bayesian sequential change diagnosis [5] studies a problem formulation where a single fault occurs but there are $M$ causes for failure. The goal is to detect when the fault happens and what caused it. Complete observability of the information is assumed. Our proposed formulation in contrast imposes observation and fault interaction structures for multiple simultaneous faults creating a completely new class of problems that cannot be mapped into this framework.

There is a sizable literature on sensor failure detection in the context of sensor networks [4], including detection of failures in multiple sensors [9], [10]. Many heuristics based on practical requirements have been proposed [14], [17], [6], [26], but none have optimality guarantees nor the change point structure is properly explored. In contrast we propose an algorithm with performance guarantees using a novel change point formulation. In fact, our analysis in this paper applies to commonly used correlation tracking heuristics (e.g. [19]) and shows that without properly structured stopping times that exchange information, these sensor fault detection heuristics can perform very poorly.

## C. Paper organization

The paper is organized as follows. Section II states the problem in more detail and establishes some basic notation. Section III investigates the delay of the natural extension procedure based on posterior probabilities. Section IV introduces STIE (Localized Stopping Time with Information Exchange) and analyzes its performance. It also calculates the best possible delay achievable by *any* procedure. Section V presents simulation examples. Section VII presents technical assumptions and proofs. Section VI concludes the paper with a discussion of the results and avenues of future work.

Parts of this work have been presented in IPSN 2008 (Information Processing for Sensor Networks) and the 2nd International Workshop on Sequential Analysis.

## II. PROBLEM STATEMENT AND NOTATION

Consider the setup given by the communication graph in Figure 2(a) and fault graph in Figure 2(b). Two subsystems 1 and 2 fail at random times $\lambda_1$ and $\lambda_2$ respectively. Subsystem 1 observes the private information sequence of random variables $X = \{X_n, n \geq 1\}$. The distribution of this sequence experiences a change due to the change time $\lambda_1$ of subsystem 1. Using our earlier notation, $f_0(X) \xrightarrow{\lambda_1} f_1(X)$, where $f_0(X)$ and $f_1(X)$ are known densities specific to random variable $X$. Similarly subsystem 2 observes the private information sequence $Y = \{Y_n, n \geq 1\}$ and its distribution follows $f_0(Y) \xrightarrow{\lambda_2} f_1(Y)$. Both subsystems observe the shared information as the random variable sequence $Z = \{Z_n, n \geq 1\}$. Its distribution changes according to $f_0(Z) \xrightarrow{\min(\lambda_1, \lambda_2)} f_1(Z)$, i.e.

$$Z_n \overset{i.i.d}{\sim} f_0(Z), \quad n < \min(\lambda_1, \lambda_2),$$
$$\overset{i.i.d.}{\sim} f_1(Z), \quad n \geq \min(\lambda_1, \lambda_2).$$

The sequence of random variables for $X$ between time $k$ and $n$ is denoted as $\mathbf{X}_n^k$, and similarly for other random variables. The goal of multiple subsystem simultaneous fault detection is to construct stopping times $\bar{\nu}_1$ for subsystem 1, only using $\{\mathbf{X}_n^1, \mathbf{Z}_n^1\}$ at time $n$, and $\bar{\nu}_2$ for subsystem 2, only using $\{\mathbf{Y}_n^1, \mathbf{Z}_n^1\}$ at time $n$, that detect whether $\lambda_1 \leq n$ and $\lambda_2 \leq n$ efficiently. Efficiency is measured according to the performance metrics in Section II-A, i.e., each stopping time achieves a small detection delay for a given false alarm. The multiple interacting fault detection problem is difficult due to the interacting nature of the faults and the information sharing pattern imposed by the communication graph. Furthermore, these constraints make it hard, if not impossible, to find an optimal stopping time in the spirit of Shiryaev [23], i.e. that is non-asymptotic optimal. Therefore, it is natural to seek stopping times that can achieve asymptotic optimality.

To conclude, we detail further the Bayesian formulation of the multiple interacting fault problem. The fault graph in Figure 2(b) details the probability dependency structure. Conditional on the change times $\lambda_1$ and $\lambda_2$, the random variables $X$, $Y$ and $Z$ are all independent. We also assume the joint prior distribution of the change times is denoted $\mathbb{P}(\lambda_1 = k_1, \lambda_2 = k_2) = \pi_1(k_1)\pi_2(k_2)$. For convenience, define the cumulative quantities $\Pi_n^1 = \mathbb{P}(\lambda_1 > n)$ and $\Pi_n^2 = \mathbb{P}(\lambda_2 > n)$.

The $\sigma$-field generated by a sequence such as $\mathbf{X}_n^1$ is denoted by $\mathcal{F}_X^n$. For the fields of joint variables, we use notation such as $\mathcal{F}_{X,Y}^n$. Based on these definitions we can formalize the restriction that subsystem 1 can only use use random variables $X$ and $Z$ for its decision, whereas subsystem 2 can only use random variables $Y$ and $Z$ for its decision, by requiring that the respective stopping rules be localized:

**Definition 1** (Localized stopping time). *A localized stopping time for subsystem 1 is a stopping time $\nu_1 \in \mathcal{F}_{X,Z}^n$. Similarly, a localized stopping time for subsystem 2 is a stopping time $\nu_2 \in \mathcal{F}_{Y,Z}^n$.*

The probability measure in the joint space of random variables when the change happens at $\lambda_1 = k_1$ and $\lambda_2 = k_2$ is defined as:

$$\begin{aligned}
\mathbb{P}_{k1,k2}(\mathbf{X}_n^1, \mathbf{Y}_n^1, \mathbf{Z}_n^1) &= \mathbb{P}_{k_1}(\mathbf{X}_n^1)\mathbb{P}_{k_1 \wedge k_2}(\mathbf{Z}_n^1)\mathbb{P}_{k_2}(\mathbf{Y}_n^1) \\
&\sim \prod_{i=1}^{k_1-1} f_0(X_i) \prod_{i=k_1}^{n} f_1(X_i) \prod_{i=1}^{k_1 \wedge k_2 - 1} f_0(Z_i) \prod_{i=k_1 \wedge k_2}^{n} f_1(Z_i) \prod_{i=1}^{k_2-1} f_0(Y_i) \prod_{i=k_2}^{n} f_1(Y_i) \\
&= \mathrm{L}_{k_1}(\mathbf{X}_n^1)\mathrm{L}_{k_1 \wedge k_2}(\mathbf{Z}_n^1)\mathrm{L}_{k_2}(\mathbf{Y}_n^1).
\end{aligned}$$

We define $\mathrm{L}_{k_1}(\mathbf{X}_n^1)$ denotes the product of densities for $X$ and similarly for other variables. From the definitions, when $\lambda_2 = \infty$ we have $\mathbb{P}_{k_1,\infty}(\mathbf{X}_n^1, \mathbf{Y}_n^1, \mathbf{Z}_n^1) = \mathbb{P}_{k_1}(\mathbf{X}_n^1)\mathbb{P}_{k_1}(\mathbf{Z}_n^1)\mathbb{P}_{\infty}(\mathbf{Y}_n^1)$. The appropriate marginalized measures are also defined, such as:

$$\mathbb{P}_{\lambda_1,\lambda_2}(\mathbf{X}_n^1, \mathbf{Y}_n^1, \mathbf{Z}_n^1) = \sum_{k_1=1}^{n}\sum_{k_2=1}^{n} \pi(k_1)\pi(k_2)\mathbb{P}_{k1,k2}(\mathbf{X}_n^1, \mathbf{Y}_n^1, \mathbf{Z}_n^1).$$

In our notation $\mathbb{E}_{k_1,k_2}$ refers to expectations with respect to the measure $\mathbb{P}_{k_1,k_2}(\mathbf{X}_n^1, \mathbf{Y}_n^1, \mathbf{Z}_n^1)$. It will be useful to define the log-likelihood ratio of sample $i$ for random variable $X_i$ and the accumulated log-likelihood:

$$r_i(X) = \log\left(\frac{f_1(X_i)}{f_0(X_i)}\right) \; ; \; R_n^k(X) = \sum_{i=k}^{n} r_i(X). \tag{6}$$

Similar definitions hold for all random variables. We make assumptions about the expectations of the log-likelihoods under pre-change and post-change distributions. In particular, assume they are all finite ($*$ denotes *don't care*):

$$\mathbb{E}_{1,*}[r_i(X)] = \int f_1(x) \log \frac{f_1(x)}{f_0(x)} \mu(dx) = D(f_1(X)\|f_0(X)) = q_1(X),$$

where $\mu$ is the Lebesgue measure. Similarly,

$$\mathbb{E}_{\infty,*}[r_i(X)] = \int f_0(x) \log \frac{f_0(x)}{f_1(x)} \mu(dx) = -D(f_0(X)\|f_1(X)) = -q_0(X).$$

For $Y$ a similar assumption holds, only noting that expectations will be with respect to $\mathbb{E}_{*,0}$ and $\mathbb{E}_{*,\infty}$. For $Z$, again the definitions hold, but expectations should be with respect to $\mathbb{E}_{0,0}$ and $\mathbb{E}_{\infty,\infty}$. The assumption is that $q_0(X), q_1(X), q_0(Z), q_1(Z), q_0(Y)$ and $q_1(Y)$ are all positive and finite.

Further detailed technical assumptions are stated in Section VII-A.

### A. Performance metrics

Denote the fault detection rule for subsystem $u$ by stopping time $\nu_u$ for $u = 1$ and $u = 2$. In the change point literature, such a stopping time is evaluated according to two metrics: probability of false alarm and detection delay, see e.g., [25].

**Definition 2** (Probability of false alarm). *Given a stopping time $\nu_u$ and the change time $\lambda_u$ the false alarm probability at $\lambda_u = k_u$ is defined as*

$$P_{fa}^{(k_1,k_2)}(\nu_u) = \mathbb{P}_{k_1,k_2}(\nu_u < k_u).$$

*The false alarm probability for procedure $\nu_u$ is given by*

$$P_{fa}^{\pi_1,\pi_2}(\nu_u) = \mathbb{P}_\lambda(\nu_u < \lambda_u) = \sum_{k_1=1}^{\infty} \sum_{k_2=1}^{\infty} \pi_1(k_1)\pi_2(k_2)\mathbb{P}_{k_1,k_2}(\nu_u < k_u).$$

*The marginal false alarm probabilities for procedures $\nu_1$ and $\nu_2$ are*

$$MP_{fa}^{(\lambda_1,k_2)}(\nu_1) = \mathbb{P}_{\lambda_1,k_2}(\nu_1 < \lambda_1) \text{ and } MP_{fa}^{(k_1,\lambda_2)}(\nu_2) = \mathbb{P}_{k_1,\lambda_2}(\nu_2 < \lambda_2).$$

*The conditional marginal false alarm probabilities for procedures $\nu_1$ and $\nu_2$ are*

$$MP_{fa}^{\pi_1,\pi_2}(\nu_1|\lambda_2 < \lambda_1) = \sum_{k_1=1}^{\infty} \pi_1(k_1) \sum_{k_2=1}^{k_1-1} \pi_2(k_2)\mathbb{P}_{\infty,k_2}(\nu_1 < k_1) \text{ and}$$

$$MP_{fa}^{\pi_1,\pi_2}(\nu_2|\lambda_1 < \lambda_2) = \sum_{k_2=1}^{\infty} \pi_2(k_2) \sum_{k_1=1}^{k_2-1} \pi_1(k_1)\mathbb{P}_{k_1,\infty}(\nu_2 < k_2).$$

**Definition 3** (Detection delay). *The m-th moment of the delay of a sequential procedure $\nu_u$ for change time $\lambda_u = k_u$ is defined as*

$$D_m^{(k_1,k_2)}(\nu_u) = \mathbb{E}_{k_1,k_2}\left[(\nu_u - k_u)^m | \nu_u \geq k_u\right].$$

*The m-th moment of the detection delay is*

$$D_m^{\pi_1,\pi_2}(\nu_u) = \mathbb{E}_{\lambda_1,\lambda_2}\left[(\nu_u - \lambda_u)^m | \nu_u \geq \lambda_u\right] = \sum_{k_1=1}^{\infty} \sum_{k_2=1}^{\infty} \pi_1(k_1)\pi_2(k_2)D_m^{(k_1,k_2)}(\nu_u).$$

A good procedure achieves small (even minimum) delay $D_m^{\pi_1,\pi_2}(\nu_u)$, while maintaining $P_{\text{fa}}^{\pi_1,\pi_2}(\nu_u) \leq \alpha$, for a pre-specified $\alpha$. An optimal detection procedure for subsystem $u$ is a procedure $\nu_u$ for which the delay $D_1^{\pi_1,\pi_2}(\nu_u)$ is minimized while keeping the false alarm below a chosen probability $\alpha$ so that $P_{\text{fa}}^{\pi_1,\pi_2}(\nu_u) \leq \alpha$. Such a rule is called an *optimal sequential procedure*. Notice that a procedure that satisfies $P_{\text{fa}}^{\pi_1,\pi_2}(\nu_u) \leq \alpha$ does not necessarily satisfy $MP_{\text{fa}}(\nu_u) \leq \alpha$, and in particular this is true for the optimal sequential procedure.

## III. LOCALIZED FAULT DETECTION WITHOUT INFORMATION EXCHANGE

One approach to solving the multiple interacting fault detection problem is to use a methodology inspired by solving a single change point problem. We first review the relevant solution for a single change point problem and then we describe the natural extension which is shown to be unable to exploit the common information available for detection of both change points.

### A. Test statistic for a single change point

Suppose a single subsystem fails at a random time $\lambda$, with distribution $\mathbb{P}(\lambda = n) = \pi_1(n)$. The observations $X$ are an identical independently distributed random variable, with distribution $f_0$ before change and $f_1$ after change. The fault detection formulation for this single subsystem is the standard single change point detection problem.

Shiryaev [24] showed that an *optimal sequential procedure* is the procedure that tests the hypothesis $H_1 : \lambda \leq n$ against $H_0 : \lambda > n$ at each $n$, using the observations $X_1, ..., X_n$. The Shiryaev-Robert-Polak (SRP) sequential procedure is a threshold test on the posterior probability as shown in Eq. (1). The SRP test quantity can be further developed as

$$\Lambda_n(X) = \frac{\mathbb{P}(\lambda \leq n \,|\, \mathcal{F}_X^n)}{1 - \mathbb{P}(\lambda \leq n \,|\, \mathcal{F}_X^n)} = \frac{\sum_{k=0}^{n} \pi_1(k) \prod_{r=1}^{k} f_0(X_r) \prod_{r=k+1}^{n} f_1(X_r)}{\sum_{k=n+1}^{\infty} \pi(k) \prod_{r=1}^{n} f_0(X_r)} = \Lambda_0 + \Pi_n^{-1} \sum_{k=1}^{n} \pi_1(k) e^{R_n^k(X)},$$

where $R_n^k(X)$ is defined in Eq. (6), $\Lambda_0 = \pi_1(0)/(1-\pi_1(0))$ and $\Pi_n = \mathbb{P}(\lambda > n)$. This test quantity in the stopping time in Eq. (2) with threshold rule given by Eq. (3) to obtain the SRP procedure. Tartakovsky et al [25] showed the SRP procedure achieves the *optimal asymptotic delay* for the problem of minimizing the expected delay constrained to a false alarm probability $\alpha$ (i.e. $P_{\text{fa}}^\pi(\nu_S) \leq \alpha$). Furthermore, the asymptotic delay as $\alpha \to 0$ is given by Eq. (4), which *matches the lower bound for delays for any procedure with false alarm $\alpha$*.

The single change point problem is considerably simpler than the multiple change problem, since once a change is detected, it is attributed to a unique fault, and there is no chance of *confusion* with other potentially failed subsystems. We summarize the important facts in the following definition:

**Definition 4** (Sequential test statistic)**.** *The generalization of the test for a SRP procedure using random variables $X$ and $Z$ is*

$$\Lambda_n(X, Z) = \Lambda_0 + \Pi_n^{-1} \sum_{k=1}^{n} \pi_1(k) e^{R_n^k(X) + R_n^k(Z)}. \tag{7}$$

*This corresponds to the ratio in Eq. (5). The corresponding stopping time is $\nu_1 = \nu_s(X, Z)$ and uses the threshold in Eq. (3). Similarly we can define $\Lambda_n(X)$, $\Lambda_n(Y, Z)$ and $\Lambda_n(Y)$, using $\pi_1$, $\pi_2$, $Y$ and $Z$. The corresponding stopping times are $\tilde{\nu}_1 = \nu_s(X)$, $\nu_2 = \nu_s(Y, Z)$ and $\tilde{\nu}_2 = \nu_s(Y)$.*

**Remark 1.** *For the rest of the paper, we assume without loss of generality that $\Lambda_0 = 0$ for the SRP procedure.*

*B. Test statistic for multiple interacting change points*

In this section we study the first natural approach to solving the multiple interacting fault detection problem. We focus on subsystem 1. Heuristically, a threshold test in the posterior probability seems a reasonable choice for stopping time. For the single change point case this is an optimal choice. In the modified framework, such a choice may not be optimal, but it is certainly an attractive and simple test. Intuitively, this is the first test one would consider. The posterior probability test can be written as:

$$\begin{aligned} \nu_1(X,Z) &= \inf\{n : p_n(X,Z) \geq 1-\alpha\}, \\ p_n(X,Z) &= \mathbb{P}_{\lambda_1,\lambda_2}(\lambda_1 \leq n \mid \mathcal{F}_{X,Z}^n). \end{aligned} \qquad (8)$$

To put into standard form, notice that

$$\frac{p_n(X,Z)}{1-p_n(X,Z)} \geq \frac{1-\alpha}{\alpha},$$

is an equivalent test to the original. Then the test statistic is

$$\Lambda_n^{\text{noex}}(X,Z) = \frac{\mathbb{P}_{\lambda_1,\lambda_2}(\lambda_1 \leq n \mid \mathcal{F}_{X,Z}^n)}{\mathbb{P}_{\lambda_1,\lambda_2}(\lambda_1 > n \mid \mathcal{F}_{X,Z}^n)}.$$

From the problem definition, we can compute the probabilities involved in the statistic $\Lambda_n^{\text{noex}}(X,Z) = a_n/b_n$:

$$\begin{aligned} a_n &= \sum_{k_1=1}^{n}\sum_{k_2=1}^{\infty} \pi_1(k_1)\pi_2(k_2)\mathrm{L}_{k_1}(\mathbf{X}_n^1)\mathrm{L}_{k_1 \wedge k_2}(\mathbf{Z}_n^1), \\ b_n &= \Pi_{1,n}\mathrm{L}_{n+1}(\mathbf{X}_n^1)\left\{\Pi_{2,n}\mathrm{L}_{n+1}(\mathbf{Z}_n^1) + \sum_{k_2=1}^{n}\pi_2(k_2)\mathrm{L}_{k_2}(\mathbf{Z}_n^1)\right\}. \end{aligned}$$

Similarly, we can define a stopping time based on $\Lambda_n^{\text{noex}}(Y,Z)$ for subsystem 2. The first important observation is that computing the test quantity is non-trivial. More importantly and somewhat surprisingly, no delay reduction benefit is obtained from using the shared information:

**Theorem 1.** *Assume $q_0(Z) \geq q_1(Z)$ or $\pi_2(k_2) > 0$ for $k_2 \geq K_2$. For the posterior threshold test $\nu_1(X,Z)$ without information exchange given by Eq. (8), the delay satisfies (as $\alpha \to 0$)*

$$\begin{aligned} D_1^{k_1,k_2}(\nu_1(X,Z)) &\geq D_1^{k_1,k_2}(\nu_1(X)), \\ D_1^{\pi_1,\pi_2}(\nu_1(X,Z)) &\geq D_1^{\pi_1,\pi_2}(\nu_1(X)). \end{aligned}$$

*For the threshold test $\nu_2(Y,Z)$, similar bounds apply.*

The result shows that the performance of the rule does not depend on the statistics of the shared information $Z$. This is a surprising result, since we expect an improvement in performance of the order of the KL divergence ($q_1(Z)$) for the pre and post-change distributions of $Z$.

Thus, in this procedure the shared information is not useful in determining which subsystem failed. The information in either pair $(X,Z)$ or $(Y,Z)$ by itself is not helpful in determining whether the change in $Z$ is induced by a failure in subsystem 1 or in subsystem 2. In the hypothesis test, the null hypothesis as well as the alternative hypothesis incorporate the information that a change in the shared information could have happened because the other subsystem failed.

## IV. STIE: A LOCALIZED STOPPING TIME WITH INFORMATION EXCHANGE

In this Section we propose localized STIE (Stopping Time with Information Exchange), an interacting stopping time method that attempts to overcome the limitations discussed in the previous section and benefit from shared information.

The structure of the interaction between faults leads to an observation: before either subsystem has failed, the shared information helps both decide whether they are failed or not; after one of them fails, the shared information is not useful for the non-failed one. STIE relies on this observation by initially computing a test statistic for subsystem 1 that assumes subsystem 2 is not failed. This is just the standard SRP procedure, that uses both $X$ and $Z$ and can be computed as shown in Definition 4. A similar test statistic is computed for subsystem 2.

Based on these statistics, we can define the stopping rule $\nu_1$ for subsystem 1:

$$\nu_1 = \min\{n : \Lambda_n(X, Z) \geq B_\alpha\} = \nu_S(X, Z), \tag{9}$$

and similarly $\nu_2$ for subsystem 2. Each subsystem computes this test, until one of them believes it is failed. Say subsystem 2 believes it is failed at time $n$ (so $\nu_2 = n$) and before subsystem 1 (i.e. $\nu_1 > n$). In STIE, subsystem 2 communicates its decision to subsystem 1. Then subsystem 1 should not use the shared information anymore, else it may be misled to think it has failed due to the change in $Z$. From this point onwards, subsystem 1 computes the SRP posterior rule only based on its private information $X$ and computes the stopping rule $\tilde{\nu}_1$ until failure:

$$\tilde{\nu}_1 = \min\{n : \Lambda_n(X) \geq B_\alpha\} = \nu_S(X). \tag{10}$$

If instead subsystem 1 had declared failure first using $\nu_1$, subsystem 2 would use an analogous stopping rule $\tilde{\nu}_2$. We can summarize formally STIE in terms of composite stopping rules $\bar{\nu}_1$ for subsystem 1

$$\bar{\nu}_1 = \nu_1 \mathbb{I}(\nu_1 \leq \nu_2) + \max(\tilde{\nu}_1, \nu_2) \mathbb{I}(\nu_1 > \nu_2) \tag{11}$$

and $\bar{\nu}_2$ for subsystem 2

$$\bar{\nu}_2 = \nu_2 \mathbb{I}(\nu_2 \leq \nu_1) + \max(\tilde{\nu}_2, \nu_1) \mathbb{I}(\nu_2 > \nu_1). \tag{12}$$

In the composite rule $\bar{\nu}_1$, the exchanged bit is represented by the indicator $\mathbb{I}(\nu_1 > \nu_2)$. The $\max$ operator reflects the situation the private information from a subsystem dictates that it has already failed (e.g., $\tilde{\nu}_1 < \nu_2 = n$), in which case one should stop immediately at the present time ($\nu_2 = n$).

The proposed stopping rules can be implemented in the system in Figure 2(b) in a distributed way. In STIE there is an information exchange between subsystems, but it is constrained to a single bit that informs when a subsystem's statistic has crossed its threshold. Then the other subsystem stops using the shared information (that is, it recomputes its own statistics without using shared information). This is a new feature of the model investigated in this paper. Previous literature in distributed hypothesis testing focused in the case where all subsystems observed the same hypothesis. Here we have a problem where subsystems observe hypothesis that interact with each other.

Another important benefit of STIE is that it can be computed efficiently, as each subsystem only computes two recursions following Definition 4, without requiring the storage of all observed values of the random variables $X, Y, Z$. In [21] we describe a detailed efficient implementation in an application setting.

STIE is summarized as follows. Each subsystem computes posteriors as if the other subsystem is always working, until the time one of them declares itself as failed. Both subsystems at this point are using shared information. When one subsystem is thought to have failed the other stops using the shared information, and recomputes the change point test using only its private information.

In the remainder of the section we compute the false alarm probability and the detection delay for STIE. The detection with information exchange algorithm is interesting if we are able to show that for a given false alarm rate $O(\alpha)$, it achieves expected delays smaller than if the shared information is not used.

## A. Performance analysis: false alarm

From Definition 2, the false alarm for subsystem 1 is given by

$$P_{\text{fa}}^{\pi_1,\pi_2}(\bar{\nu}_1) = \mathbb{P}_{\lambda_1,\lambda_2}(\bar{\nu}_1 < \lambda_1). \quad (13)$$

Moreover, by design choice of the threshold for tests $\nu_1$ and $\tilde{\nu}_1$ that form STIE, the false alarm when there is no change observed in subsystem 2 (i.e. $\lambda_2 = \infty$) is bounded:

$$P_{\text{fa}}^{\pi_1,\infty}(\nu_1) = \mathbb{P}_{\lambda_1,\infty}(\nu_1 < \lambda_1) \le \alpha,$$
$$P_{\text{fa}}^{\pi_1,\infty}(\tilde{\nu}_1) = \mathbb{P}_{\lambda_1,\infty}(\tilde{\nu}_1 < \lambda_1) \le \alpha.$$

Unfortunately these guarantees do not translate into a guarantee for the false alarm in Eq. (13) of the procedure composed of both tests. Analyzing Eqns. (11) and (12) we notice that subsystem 1 can raise two kinds of false alarm at some time $n$: one caused without any change ($\lambda_1 > n$ and $\lambda_2 > n$), and another caused when the shared information experiences a change due to a fault in subsystem 2 ($\lambda_2 \le n$). Based on this observation we define the error coupling probability:

**Definition 5** (Error decoupling probability). *The error decoupling probabilities of stopping times in a set of procedures $(\bar{\nu}_1, \bar{\nu}_2)$ are defined as*

$$\xi_{\lambda_1,\lambda_2}^{\alpha}(\bar{\nu}_1) = \mathbb{P}_{\lambda_1,\lambda_2}(\bar{\nu}_1 \le \bar{\nu}_2, \lambda_2 \le \bar{\nu}_1 < \lambda_1),$$
$$\xi_{\lambda_1,\lambda_2}^{\alpha}(\bar{\nu}_2) = \mathbb{P}_{\lambda_1,\lambda_2}(\bar{\nu}_2 \le \bar{\nu}_1, \lambda_1 \le \bar{\nu}_2 < \lambda_2).$$

*A **regular** fault detection procedure is a set of procedures for which the following conditions hold:*

$$\lim_{\alpha \to 0} \xi_{\lambda_1,\lambda_2}^{\alpha}(\bar{\nu}_1) = 0,$$
$$\lim_{\alpha \to 0} \xi_{\lambda_1,\lambda_2}^{\alpha}(\bar{\nu}_2) = 0.$$

*A **strong** fault detection procedure is a set of procedures that has*

$$\xi_{\lambda_1,\lambda_2}^{\alpha}(\bar{\nu}_1) = O(\alpha),$$
$$\xi_{\lambda_1,\lambda_2}^{\alpha}(\bar{\nu}_2) = O(\alpha).$$

The importance of this definition is shown in Theorem 2: the false alarm for the composite procedure STIE can be shown to be bounded by the sum of the desired false alarm rate ($\alpha$) and the error coupling probability. This probability measures the degree of coupling caused by the competing change time. If it is of order $O(\alpha)$, we say the procedure is regular. As a comparison, if the event $\mathcal{E} = \{\bar{\nu}_1 < \lambda_1\}$ was contained in the union of the events $\mathcal{E}_1 = \{\nu_1 < \lambda_1, \lambda_2 = \infty\}$ and $\mathcal{E}_2 = \{\tilde{\nu}_1 < \lambda_1, \lambda_2 = \infty\}$, then a direct union bound shows the false alarm to be bounded by $2\alpha$.

**Theorem 2** (False alarm of STIE).

(a) *The probability of false alarm of subsystem 1 for the joint procedure with information exchange (STIE) is bounded by:*

$$P_{fa}^{\pi_1,\pi_2}(\bar{\nu}_1) \leq \alpha + \xi_{\lambda_1,\lambda_2}^{\alpha}(\bar{\nu}_1).$$

(b) *The marginal probability of false alarm of subsystem 1 for STIE is bounded by:*

$$MP_{fa}^{\pi_1,k_2}(\bar{\nu}_1) \leq \alpha + \xi_{\lambda_1,k_2}^{\alpha}(\bar{\nu}_1).$$

The intuition behind Theorem 2 is that if the decision of not using $Z$ was immediate, as soon as $\min(\lambda_1, \lambda_2)$ happens, there would be no error coupling event and the composite procedure would have false alarm $\alpha$. But due to delayed detection, there is a period of time when subsystem 1 can declare a fault due to change only in $Z$ but not in $X$. This is exactly $\mathcal{E}_c = \{\lambda_2 < \bar{\nu}_1 < \lambda_1, \bar{\nu}_1 \leq \bar{\nu}_2\}$, the error event coupling the tests between subsystems. If the asymptotic rate of $\xi$ with $\alpha$ is faster than $O(\alpha)$ (procedure is strong), the additional false alarm incurred is not significant, since delay is proportional to the logarithm of false alarm. Otherwise, the error incurred is significant, and reduces any potential delay benefits. Theorem 3 completes the understanding about the false alarm for procedure STIE by analyzing the error coupling probability and identifying under what conditions the procedure is strong, and therefore has false alarm rate of order $\alpha$.

**Theorem 3** (Error coupling probability regularity). *The theorem is stated for subsystem 1. For subsystem 2 it suffices to exchange the role of $X$ and $Y$.*

(a) *The procedure STIE is a regular fault detection procedure.*

(b) *Let assumptions VII.2 and VII.5 hold. Define $b_1 = q_0(X) - q_1(Z) + d_1$ and the rate*

$$r_a^* = \frac{1}{w^*} \frac{[\min\{q_0(X), q_1(Z)\} + q_1(Y) + d_1 - d_2]^2}{\max\{\sigma_0^2(X), \sigma_1^2(Z)\} + \sigma_1^2(Y)},$$

*where*

$$w^* = \sqrt{\frac{\sigma_1^2(X) + \sigma_1^2(Z)}{\max\{\sigma_0^2(X), \sigma_1^2(Z)\} + \sigma_1^2(Y)}} [\min\{q_0(X), q_1(Z)\} + q_1(Y) + d_1 - d_2] - b_1,$$

*constants $\sigma_0^2(X)$, $\sigma_1^2(Z)$ and $\sigma_1^2(Y)$ are defined in assumption VII.2 and constants $d_1$ and $d_2$ are defined in assumption VII.1. Then*

$$\lim_{\alpha \to 0} \frac{\log \xi_{\lambda_1,\lambda_2}^{\alpha}(\bar{\nu}_1)}{\log \alpha} \geq r^*,$$

*where*

(a) *If $b_1 \leq 0$ then $r^* = r_a^*$;*
(b) *If $b_1 > 0$ then $r^* = \max(r_a^*, r_b^*)$, where*

$$r_b^* = 4 \frac{b_1}{\sigma_1^2(X) + \sigma_1^2(Z)}.$$

*Therefore, if $r^* > 1$, STIE is a strong fault detection procedure.*

The main element to proving this theorem is an identification of which types events cause strong error coupling. An important key result (Lemma 1) shows that for STIE, conditions on the amount of information provided by the different information sets suffice for the various types of errors to be decoupled.

**Example.** Let us consider a simple scenario where $\sigma_0^2(X) = \sigma_1^2(Z) = \sigma_1^2(Y) = 1/2$ and $d_1 = d_2 = \epsilon$, where $\epsilon$ is small. When the shared information is stronger than the private information (i.e., $q_0(X) < q_1(Z)$), and if $q_0(X)$ is small, then $b_1 < 0$ and $r^* \approx q_1(Y)^2/(q_1(Y) + q_1(Z))$. So the private information of subsystem 2 needs to be large as well (i.e., $q_1(Y) = O(\sqrt{q_1(Z)})$ for the procedure to be strong. Intuitively, this will prevent subsystem 1 being misled by a fault in subsystem 2, since subsystem 2 quickly detects its own fault. Otherwise, if (i.e., $q_0(X) >> q_1(Z)$), then $r^* \geq r_b^*$ and $r_b^* \approx 4q_0(X)$, so the procedure is strong if sufficient private information is available to subsystem 1, independent of subsystem 2. If $q_0(X)$ is small in this case, then the procedure still benefits from the strength of private information of subsystem 2.

In more general scenarios, the error coupling events can be inferred from the fault graph structure. Then, if the probability of error coupling is small, the inference problem for each subsystem can be decoupled, and thus behaves as multiple single fault problems. A procedure like STIE is then strong if both private and shared information ($X$, $Y$ or $Z$) are relatively too strong.

### B. Performance analysis: detection delay

The performance of individual procedures that compose STIE are known under the condition no change occurs in the competing subsystem. For example, for subsystem 1 if $\lambda_2 = \infty$, it is clear that the standard delay computation in Eq. (4) applies to stopping rules $\nu_1$ and $\tilde{\nu}_1$ with appropriately chosen constants. We can define the detection delay constants for each individual change point that composes STIE:

**Definition 6** (Detection delays). *Define the following detection delay constants:*

$$L_1^\alpha = \frac{|\log \alpha|}{q_1(X) + q_1(Z) + d_1}, \quad L_2^\alpha = \frac{|\log \alpha|}{q_1(Y) + q_1(Z) + d_2},$$

$$\tilde{L}_1^\alpha = \frac{|\log \alpha|}{q_1(X) + d_1}, \quad \tilde{L}_2^\alpha = \frac{|\log \alpha|}{q_1(Y) + d_2},$$

*where $d_1$ is the rate for prior $\pi_1$ and $d_2$ is the rate for prior $\pi_2$ according to Assumption VII.1.*

Based on this definition and under the condition $\lambda_2 = \infty$, notice that $D_1^{\lambda_1,\lambda_2}(\nu_1) \doteq L_1^\alpha$ and $D_1^{\lambda_1,\lambda_2}(\tilde{\nu}_1) \doteq \tilde{L}_1^\alpha$. Furthermore, $D_1^{\lambda_1,\lambda_2}(\nu_1) < D_1^{\lambda_1,\lambda_2}(\tilde{\nu}_1)$. $L_1^\alpha$ is the smallest delay achievable in this problem as it assumes the shared information only changes due to $\lambda_1$. $\tilde{L}_1^\alpha$ is the delay achieved by $\tilde{\nu}_1$ in the general scenario since it does not use the shared information. Thus, we expect $D_1^{\lambda_1,\lambda_2}(\nu_1) \leq D_1^{\lambda_1,\lambda_2}(\bar{\nu}_1) \leq D_1^{\lambda_1,\lambda_2}(\tilde{\nu}_1)$ as when $\lambda_2$ happens much earlier than $\lambda_1$, $STIE$ will use $\tilde{\nu}_1$ for subsystem 1. This might even be the case for any procedure respecting the false alarm bound.

It is natural to start the analysis by determining an asymptotic lower bound for the detection delay of *any* procedure for multiple interacting fault detection. The minimization is constrained by the desire for the procedure to incur false alarm at most $\alpha$. For this to hold we consider only procedures in an appropriate false alarm class:

**Definition 7** (False alarm classes). *For stopping times $\nu_1(X, Z)$ dependent only on $X$ and $Z$ define the classes:*

(i) $\Delta_1(\alpha)$ *such that* $P_{fa}^{\pi_1, \infty}(\nu_1) \leq \alpha$,
(ii) $\tilde{\Delta}_1(\alpha, k_2)$ *such that* $MP_{fa}^{\pi_1, k_2}(\nu_1) \leq \alpha$,
(iii) $\tilde{\Delta}_1(\alpha)$ *such that* $MP_{fa}^{\pi_1, \tilde{\pi}_2}(\nu_1 | \lambda_2 < \lambda_1) \leq \alpha$.

*Also, define similar classes for stopping times $\nu_2(Y, Z)$ dependent on $Y$ and $Z$.*

Using the definition we can prove a performance lower bound for our problem among certain classes of procedures as shown in Theorem 4. The lower bound guarantees that no procedure that belongs in the given class can have delay smaller than stated. It gives us a certificate against which to check the optimality of a given procedure.

**Theorem 4** (Delay lower bound). *Let Assumptions VII.1 and VII.3. Denote $c_d = (1 + o(1))$ and consider the classes in Definition 7. Then for subsystem 1 as $\alpha \to 0$:*

$$\inf_{\nu_1 \in \tilde{\Delta}_1(\alpha, k_2)} \mathbb{E}_{k_1, k_2}[(\nu_1 - k_1)^m | \nu_1 \geq k_1] \geq \left[ (L_1^\alpha)^m \mathbb{I}(k_1 \leq k_2) + (\tilde{L}_1^\alpha)^m \mathbb{I}(k_1 > k_2) \right] c_d,$$

$$\inf_{\nu_1 \in \Delta_1(\alpha) \cap \tilde{\Delta}_1(\alpha)} \mathbb{E}_{\lambda_1, \lambda_2}[(\nu_1 - \lambda_1)^m | \nu_1 \geq \lambda_1] \geq \left[ (L_1^\alpha)^m \mathbb{P}(\lambda_1 \leq \lambda_2) + (\tilde{L}_1^\alpha)^m \mathbb{P}(\lambda_1 > \lambda_2) \right] c_d.$$

*A similar result holds for subsystem 2.*

The lower bound can be intuitively understood since when $\lambda_2 < \lambda_1$, the shared information does not help in identifying the change in subsystem 1 for arbitrarily small false alarm $\alpha$. Notice the procedure $\bar{\nu}_1$ in STIE may not belong to $\Delta_1(\alpha)$, since the bound for the false alarm rate is greater than $\alpha$ and more importantly they depend on all three $X, Z$ and $Y$ by definition. But $\nu_1$ and $\tilde{\nu}_1$ do belong to $\Delta_1(\alpha)$ and $\nu_2$ and $\tilde{\nu}_2$ belong to $\Delta_2(\alpha)$.

We conclude the section computing the asymptotic delay of the procedure STIE. The main challenge in this analysis is to account for the various possible combinations of change times $\lambda_1$ and $\lambda_2$ generating different choices in the composite procedure STIE. Theorem 5 computes the detection delay of STIE under this general setup.

**Theorem 5** (Performance of STIE). *Let Assumptions VII.1 and VII.4. Consider the procedure STIE represented as the set of stopping times $(\bar{\nu}_1, \bar{\nu}_2)$. The delay of STIE as $\alpha \to 0$ is given by:*

$$D_m^{\pi_1, \pi_2}(\bar{\nu}_1) = \left[ D_m^{\pi_1, \infty}(\nu_1)(1 - \delta_\alpha) + D_m^{\pi_1, \infty}(\tilde{\nu}_1) \delta_\alpha \right] (1 + o(1)),$$

*where $\delta_\alpha = \mathbb{P}_{\lambda_1, \lambda_2}(\nu_1 > \nu_2)$, $D_m^{\pi_1, \infty}(\nu_1) \doteq (L_1^\alpha)^m$, $D_1^{\pi_1, \infty}(\tilde{\nu}_1) \doteq (\tilde{L}_1^\alpha)^m$, and $d_1$ is given in assumption VII.1. The results are also valid for $\lambda_1$ and $\lambda_2$ replaced by $k_1$ and $k_2$. For subsystem 2, analogous results apply.*

The proof of the theorem relies on careful use of concentration arguments and the fact STIE is a *regular* procedure. Notice that the asymptotic performance differs from the lower bound only on the factor $\delta_\alpha$.

**Remark 2.** *It is easy to see that in a symmetric problem (i.e. $q_i(X) = q_i(Y)$ and $\pi_1 = \pi_2$), $\delta_\alpha = \mathbb{P}(\lambda_1 \leq \lambda_2) = 1/2$, and therefore the proposed procedure achieves the delay lower bound, albeit with a potentially larger false alarm. In fact, under the conditions the procedure STIE is strong, it is actually an* optimal asymptotic procedure for the symmetric problem, *since the false alarm is bounded by $\alpha$. We conjecture this is not the case for more general scenarios, and $\delta_\alpha$ perhaps will depend on the difference between the delays of $\nu_1$ and $\nu_2$.*

## V. EXAMPLES

We evaluate the performance of our algorithm in simulations, which allows us to precisely specify the moment of failure. For the system in Figure 2 assume that $f_0(X) \sim \mathcal{N}(\mu(X), \sigma^2(X))$ and $f_1(X) \sim$ $\mathcal{N}(0, \sigma^2(X))$. Similarly, we make definitions for $Y$ and $Z$. In this case, the information strength for the private information $X$ is given by

$$q_1(X) = q_0(X) = \frac{\mu(X)^2}{\sigma^2(X)},$$

and similarly for the other information sets. Using the results obtained in Section IV-A, we can conclude that STIE is a strong fault detection procedure if

$$4\frac{q_0(X) - q_1(Z)}{2\sigma^2(X, Z)} > 1 \quad \text{and} \quad 4\frac{q_0(Y) - q_1(Z)}{2\sigma^2(Y, Z)} > 1,$$

whenever $q_1(X) > q_1(Z)$ and $q_1(Y) > q_1(Z)$. Let us assume $\mu(X) = \mu(Y) = \mu(Z) = 1$ to normalize the simulation variables. $\sigma^2(X, Z)$ is the variance of the log-likelihood under the after change measure for $Z$ and pre-change measure for $X$, which can be computed as

$$\sigma^2(X, Z) = \frac{1}{\sigma^2(X)} + \frac{1}{\sigma^2(Z)},$$

obtaining the conditions

$$\sigma^2(X) < \frac{1}{3}\sigma^2(Z) \quad \text{and} \quad \sigma^2(Y) < \frac{1}{3}\sigma^2(Z).$$

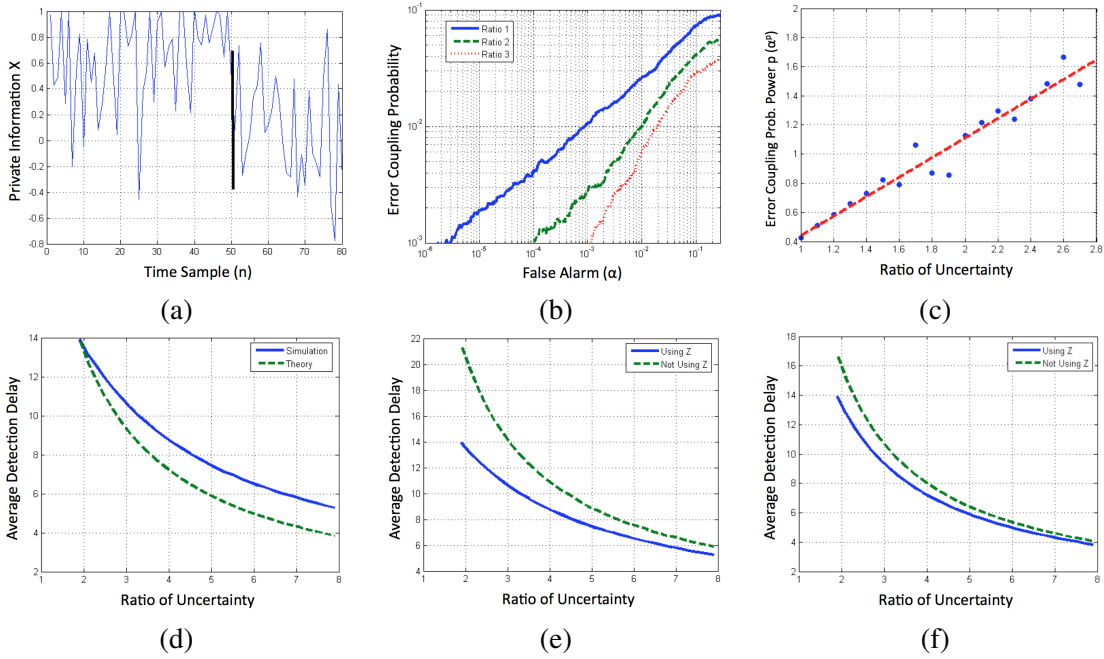

Fig. 3. Simulation example: (a)Sample path for correlation with change point at $n = 50$, (b) Error coupling probability estimates for different variance ratios and (c) Error coupling probability exponent estimates. Average delay comparison for false alarm $\alpha = 10^{-7}$ between (d) theory and simulation; (e) simulation including and excluding shared information $Z$ and (f) theory including and excluding shared information. Uncertainty ratio in these figures refers to the quantity $\sigma_Z^2/\sigma_S^2$, where $\sigma_X^2 = \sigma_Y^2 = \sigma_S^2$.

The result can be interpreted intuitively if we consider that private information $X$ focuses in capturing the behavior of $\lambda_1$ change time for subsystem 1 and similarly for $Y$ and subsystem 2. This information sets are not coupling $\lambda_1$ and $\lambda_2$. Thus, the condition implies that the information strength of these sources has to be at least three times the information strength of the shared information to avoid the coupling probability becoming too large.

For the numerical simulation, we further assume that random sequences $X$ and $Y$ are i.i.d. with variance $\sigma_S^2$. The shared information $Z$ has a fixed variance $\sigma_Z^2 = 1$. The priors for $\lambda_1$ and $\lambda_2$ are exponential distributions with rate $d_1 = d_2 = -\log(0.01)$. Figure 3(a) shows a typical sample path of private information when $\sigma_S^2 = 0.2$. Notice that without time averaging it is very hard to say exactly when the change (failure) occurred.

In Section IV-A we argued that the error coupling probability should go to zero as the false alarm rate $\alpha \to 0$ for the procedure to be consistent, and we see this in Figure 3(b). Notice though that the rate depends on the uncertainty in private information variance $\sigma_S^2$. From Figure 3(c), if $\sigma_Z^2/\sigma_S^2 < 1.8$, the error coupling probability is $O(\alpha^p)$ with $p < 1$, so the total false alarm rate of the procedure grows slower than $\alpha$. But for higher ratios, our procedure has false alarm rate $\alpha$ since the false alarm is the sum of $\alpha$ and the error coupling probability. To achieve higher ratios it is valuable to increase the private information strength, i.e., the strength of information that responds to a single fault. The theoretical prediction guarantees that the procedure is strong for ratio $\sigma_Z^2/\sigma_S^2 > 3$.

Figure 3(d) shows the theoretical and experimental average delays obtained when the threshold is $\alpha = 10^{-7}$. There is disagreement between the curves, although the qualitative behavior is as expected. The disagreement is because our results are for $\alpha \to 0$. This discrepancy is well known in sequential analysis [25]. Figure 3(e) compares the behavior of our procedure using the shared information $Z$ and one that does not use it at all. There is a substantial reduction in delay using shared information. Figure 3(f) is the corresponding theoretical prediction. There is a qualitative agreement between theory and simulation experiment.

## VI. Discussion

In this paper we developed a procedure for the multiple interacting fault detection problem. We proposed a set of basic assumptions and a framework based on the notion of a *fault graph* together with fundamental metrics to evaluate the performance of any sequential fault detection procedure. Then we proceeded to analyze the efficient algorithm STIE that achieves a good performance under the proposed metrics, and even an optimal performance under certain scenarios. As far as we know, this is the first derivation of bounds on detection delay subject to false alarm constraints in a multiple fault or multiple change point setting.

One of the main contributions of the paper is to develop a model that includes simultaneous change points that interact to generate changes in the observations. Such interactive aspect is novel and leads to a rich set of models that extend single change point modeling. Furthermore, the constraint in the information exchanged between the various tests leads to sequential tests for simultaneous hypothesis that use inconsistent views of the probability distributions.

The proposed statistical model and algorithms introduce many new ingredients into the detection literature. Due to the simultaneous and interacting change points, we develop careful asymptotic stopping time comparison calculations. Moreover, the proposed stopping times are allowed to exchange information via a network, and influence each other's behavior. This information sharing introduces coupling of the false alarm error between the procedures, and we contend that networked procedures will work well when the coupling event has a small probability. The main advantage of following such approach is that the analysis of the *decoupled* problem can benefit from the many tools developed for single change points, and care must be taken for the analysis in the coupled regime.

In the context of detection of faults in sensor networks, our algorithm performs an implicit averaging of the history of observations reducing the detection delay for a fixed false alarm. Many proposed practical methods in the literature do not perform this averaging, and therefore are subject to longer delays. Our algorithm and framework are general enough that even model based methods for computing scores, such as the one proposed in [26] or the primitive in [8], can benefit from the proposed procedure. Compared to procedures such as in [6], [17], our method benefits from implicit averaging, whereas those methods make sequential decisions based on only the current observation.

Important questions for future work are proposing and analyzing an algorithm for a general fault graphs and general communication graphs. The current framework seems to naturally lead to problems of detecting functionals of various change points observing variables whose distributions depend on these functionals. Moreover, it will also be interesting to analyze the behavior of STIE in more general graph settings. We have successfully applied the algorithm in practice [21] to more general instances of the multiple interacting fault detection problem.

## VII. Proofs

### A. Technical Assumptions

Some technical assumptions are required in order to obtain performance estimates for the procedures proposed. The first assumption is that priors have tail bounds.

**Assumption VII.1.** *The priors $\pi_1$ and $\pi_2$ of subsystems 1 and 2 satisfy the tail limit:*

$$\lim_{k_1 \to \infty} \frac{\Pi^1_{k_1+1}}{k_1} = -d_1,$$

$$\lim_{k_2 \to \infty} \frac{\Pi^2_{k_2+1}}{k_2} = -d_2.$$

The next assumption is on the tails of the log-likelihood random variables.

**Assumption VII.2.** *Assume log likelihood ratios are independent and have finite first and second moment. Denote the variance of the likelihood ratio of $X$ under $f_0$ by $\sigma_0^2(X)$ and under $f_1$ by $\sigma_1^2(X)$, of $Y$ by $\sigma_0^2(Y)$ and $\sigma_1^2(Y)$ and of $Z$ by $\sigma_0^2(Z)$ and $\sigma_1^2(Z)$. For concreteness, consider the likelihood ratio for $X$, $R_n^r(X)$. Then we assume the following tail bounds exist for $x > \mu_n^r(X)$,*

$$\mathbb{P}_{k_1,k_2}\left(R_n^r(X) > x\right) \leq K(X) \exp{-\frac{(x - \mu_n^r(X))^2}{\sigma_n^r(X)^2}}$$

*where*

$$\mu_n^r(X) = (n - k_1 \vee r)q_1(X) - (k_1 - k_1 \wedge r)q_0(X),$$
$$\sigma_n^r(X)^2 = \gamma(X)\{(n - k_1 \vee r)\sigma_1^2(X) + (k_1 - k_1 \wedge r)\sigma_0^2(X)\}.$$

*Similar bounds hold for $Y$ and $Z$, with $\mu$ and $\sigma$ appropriately defined. Also, we assume the bounds for sums, such as $R_n^r(X) + R_n^r(Z)$, by again using the appropriate definitions, such as $\mu_n^r(X,Z) = \mu_n^r(X) + \mu_n^r(Z)$ and $\sigma_n^r(X,Z)^2 = \sigma_n^r(X) + \sigma_n^r(Z)$. The constants for the bounds are defined as $K(X,Y)$ and $\gamma(X,Z)$.*

**Remark 3.** *The tail bound assumption is not overly restrictive. In fact, it only imposes a light tail constraint on the individual likelihood random variables, and then uses independence. For example, if $f_0$* and $f_1$ are Gaussian densities, the tail bounds can be obtained from large deviations. If all log likelihood ratios are bounded within interval $[-M, M]$, then using Hoeffding's bound [7], we can obtain a similar bound for each random variable, except that in this case $\gamma(X) = 2$ and

$$\sigma_n^r(X)^2 = 2\{(n - k_1 \vee r)\sigma_1^2(X) + (k_1 - k_1 \wedge r)\sigma_0^2(X) + M/3\}.$$

*The rationale behind these assumptions is that it allows precise computation of the probability of deviations of the likelihood ratio sequence, including when the maximum crosses a threshold.*

We then assume different forms of expectation concentration of the log-likelihood[12], [25].

**Assumption VII.3.** *For all $\epsilon > 0$ and $k_1, k_2 \geq 1$, as $N \to \infty$:*

$$\mathbb{P}_{k_1,k_2}\left(\frac{1}{N}\max_{1 \leq n \leq N} R_{k_1+n}^{k_1}(X) > (1+\epsilon)q_1(X)\right) \to 0,$$

$$\mathbb{P}_{k_1,k_2}\left(\frac{1}{N}\max_{1 \leq n \leq N} R_{k_1 \wedge k_2+n}^{k_1 \wedge k_2}(Z) > (1+\epsilon)q_1(Z)\right) \to 0,$$

$$\mathbb{P}_{k_1,k_2}\left(\frac{1}{N}\max_{1 \leq n \leq N} R_{k_2+n}^{k_2}(Y) > (1+\epsilon)q_1(Y)\right) \to 0.$$

**Assumption VII.4** (r-quick convergence of LLR). *The log-likelihood ratios $R_{k_1+n-1}^{k_1}(X)$, $R_{k_1 \wedge k_2+n-1}^{k_1 \wedge k_2}(Z)$ and $R_{k_2+n-1}^{k_2}(Y)$ define the stopping times:*

$$T_\epsilon^{(k_1,k_2)}(X) = \sup\left\{n \geq 1 : \left|\frac{1}{n}R_{k_1+n-1}^{k_1}(X) - q_1(X)\right| > \epsilon\right\},$$

$$T_\epsilon^{(k_1,k_2)}(Y) = \sup\left\{n \geq 1 : \left|\frac{1}{n}R_{k_1+n-1}^{k_1}(Y) - q_1(Y)\right| > \epsilon\right\},$$

$$T_\epsilon^{(k_1,k_2)}(Z) = \sup\left\{n \geq 1 : \left|\frac{1}{n}R_{k_1 \wedge k_2+n-1}^{k_1 \wedge k_2}(Z) - q_1(Z)\right| > \epsilon\right\}.$$

*For all $\epsilon > 0$ and $k_1 \geq 1$ and $k_2 \geq 1$, for some $r \geq 1$:*

$$\mathbb{E}_{k_1,k_2}\left[T_\epsilon^{(k_1,k_2)}(X)\right]^r < \infty, \quad \mathbb{E}_{k_1,k_2}\left[T_\epsilon^{(k_1,k_2)}(Y)\right]^r < \infty, \quad \mathbb{E}_{k_1,k_2}\left[T_\epsilon^{(k_1,k_2)}(Z)\right]^r < \infty,$$

$$\mathbb{E}_{\lambda_1,\lambda_2}\left[T_\epsilon^{(k_1,k_2)}(X)\right]^r < \infty, \quad \mathbb{E}_{\lambda_1,\lambda_2}\left[T_\epsilon^{(k_1,k_2)}(Y)\right]^r < \infty, \quad \mathbb{E}_{\lambda_1,\lambda_2}\left[T_\epsilon^{(k_1,k_2)}(Z)\right]^r < \infty.$$

**Assumption VII.5.** *Let*

$$S_n^{k_1}(X) := \log \frac{\pi_1(k_1)}{\Pi_1(n)} + R_n^{k_1}(X) + R_n^{k_1}(Z),$$

$$S_n^{k_2}(Y) := \log \frac{\pi_2(k_2)}{\Pi_2(n)} + R_n^{k_2}(Y) + R_n^{k_2}(Z).$$

*Let $\eta_1 = \min\{n : S_n^{k_1}(X) \geq \log B_\alpha\}$ (where $B_\alpha$ is given by Eq. (3)), and define for arbitrary $\epsilon > 0$,*

$$T_\epsilon^{k_1} = \sup\{n : |(n - k_1 + 1)^{-1}S_n^{k_1}(X) - (q_1(X) + q_1(Z) + d_1)| \geq \epsilon\}.$$

*Assume that $\mathbb{E}_{\infty,k_1} \exp T_\epsilon^{k_1} < \infty$ for any $\epsilon > 0$ and for any $k_1$. Similarly, let $\eta_2 = \min\{n : S_n^{k_2}(Y) \geq \log B_\alpha\}$, and define for arbitrary $\epsilon > 0$,*

$$T_\epsilon^{k_2} = \sup\{n : |(n - k_2 + 1)^{-1}S_n^{k_2}(Y) - (q_1(Y) + q_1(Z) + d_2)| \geq \epsilon\}.$$

*Assume that $\mathbb{E}_{\infty,k_2} \exp T_\epsilon^{k_2} < \infty$ for any $\epsilon > 0$ and for any $k_2$.*

## B. Proof of Theorem 1

We start the proof by defining an upper bound to the test statistic $\Lambda_n^{\text{noex}}(X, Z)$ that defines the stopping time $\nu_1(X, Z)$. Selecting $\bar{k}_2 = k_1 \wedge k_2$, using the assumption $\pi_2(\bar{k}_2) > 0$, we can lower bound:

$$b_n \geq \Pi_{1,n} \pi_2(\bar{k}_2) \, \mathrm{L}_{n+1}(\mathbf{X}_n^1) \mathrm{L}_{\bar{k}_2}(\mathbf{Z}_n^1),$$

so that simple algebra shows

$$\Lambda_n^{\text{noex}}(X, Z) = \frac{a_n}{b_n} \leq \Pi_{1,n}^{-1} \pi_2(\bar{k}_2)^{-1} \sum_{k_1=1}^{n} \sum_{k_2=1}^{\infty} \pi_1(k_1) \pi_2(k_2) \frac{\mathrm{L}_{k_1}(\mathbf{X}_n^1)}{\mathrm{L}_{n+1}(\mathbf{X}_n^1)}.$$

Now we can proceed as

$$\log \Lambda_n^{\text{noex}}(X, Z) = \log \frac{a_n}{b_n} \leq -\log \Pi_{1,n} - \log \pi_2(\bar{k}_2) + \log \sum_{k_1=1}^{n} \sum_{k_2=1}^{\infty} \pi_1(k_1) \pi_2(k_2) \frac{\mathrm{L}_{k_1}(\mathbf{X}_n^1)}{\mathrm{L}_{n+1}(\mathbf{X}_n^1)}$$

$$= \underbrace{-\log \pi_2(\bar{k}_2)}_{} \underbrace{- \log \Pi_{1,n} + \log \sum_{k_1=1}^{n} \pi_1(k_1) \frac{\mathrm{L}_{k_1}(\mathbf{X}_n^1)}{\mathrm{L}_{n+1}(\mathbf{X}_n^1)}}_{}$$

$$= \quad r \quad + \quad \log \Lambda_n(X),$$

where the original test statistic can be upper bounded by the sum of the standard Shryaev test for $X$ with change point at $\lambda_1$ and a positive constant $r$. Define the stopping time $\eta$

$$\eta = \inf\left\{n : \log \Lambda_n(X) + r \geq \log \frac{1-\alpha}{\alpha}\right\} \leq \nu_1(X) = \inf\left\{n : \log \Lambda_n(X) \geq \log \frac{1-\alpha}{\alpha}\right\}.$$

It is simple to see that $\eta - \nu_1(X) \to 0$ as $\alpha \to 0$. Since $\nu_1(X, Z) \geq \eta$ w.p.1, we have shown $D_1^{k_1,k_2}(\nu_1(X,Z)) \geq D_1^{k_1,k_2}(\nu_1(X))$. For $D_1^{\pi_1,\pi_2}(\nu_1(X,Z)) \geq D_1^{\pi_1,\pi_2}(\nu_1(X))$, we have the following chain of inequalities using the definition of the delays:

$$\mathbb{E}_{k_1,k_2}\left[(\nu_1(X,Z) - k_1)^m | \nu_1(X,Z) \geq k_1\right] = \frac{\mathbb{E}_{k_1,k_2}\left[[(\nu_1(X,Z) - k_1)^+]^m\right]}{\mathbb{P}_{k_1,k_2}(\nu_1(X,Z) \geq k_1)}$$

$$\geq \mathbb{E}_{k_1,k_2}\left[[(\nu_1(X,Z) - k_1)^+]^m\right]$$

$$\geq \mathbb{E}_{k_1,k_2}\left[[(\eta - k_1)^+]^m\right]$$

$$\geq (L_{\alpha,\epsilon})^m \, \mathbb{P}_{k_1,k_2}(\eta \geq k_1 + L_{\alpha,\epsilon})$$

$$\geq (L_{\alpha,\epsilon})^m \, \mathbb{P}_{k_1,k_2}(\nu_1(X) \geq k_1 + L_{\alpha,\epsilon}).$$

where $L_{\alpha,\epsilon} = (1-\epsilon)\frac{-\log e^r \alpha}{q_1(X)+d}$ and in the last line we used the Markov inequality. Lemma 4(i) also states $\mathbb{P}_{\lambda_1,\lambda_2}(\nu_1(X) \geq \lambda_1 + L_{\alpha,\epsilon}) \to 1$. Noticing this is just the expectation of the last inequality, we conclude the proof.

## C. Proof of Theorem 2

**(a)** First we show item (a) for subsystem 1. The analysis is analogous for subsystem 2.

$$\mathbb{P}_{\lambda_1,\lambda_2}(\bar{\nu}_1 < \lambda_1) = \mathbb{P}_{\lambda_1,\lambda_2}(\bar{\nu}_1 < \lambda_1, \bar{\nu}_1 > \bar{\nu}_2) + \mathbb{P}_{\lambda_1,\lambda_2}(\bar{\nu}_1 < \lambda_1, \bar{\nu}_1 \leq \bar{\nu}_2) \tag{14}$$

$$= \mathbb{P}_{\lambda_1,\lambda_2}(\max(\tilde{\nu}_1, \nu_2) < \lambda_1, \nu_1 > \nu_2) + \mathbb{P}_{\lambda_1,\lambda_2}(\bar{\nu}_1 < \lambda_1, \bar{\nu}_1 \leq \bar{\nu}_2) \tag{15}$$

$$\leq \mathbb{P}_{\lambda_1,\lambda_2}(\tilde{\nu}_1 < \lambda_1, \nu_1 > \nu_2) + \mathbb{P}_{\lambda_1,\lambda_2}(\bar{\nu}_1 < \lambda_1, \bar{\nu}_1 \leq \bar{\nu}_2) \tag{16}$$

$$\leq \alpha \mathbb{P}_{\lambda_1,\lambda_2}(\nu_1 > \nu_2) + \mathbb{P}_{\lambda_1,\lambda_2}(\bar{\nu}_1 < \lambda_1, \bar{\nu}_1 \leq \bar{\nu}_2) \tag{17}$$

$$= \alpha \mathbb{P}_{\lambda_1,\lambda_2}(\nu_1 > \nu_2) + \mathbb{P}_{\lambda_1,\lambda_2}(\bar{\nu}_1 < \lambda_1, \bar{\nu}_1 < \lambda_2, \bar{\nu}_1 \leq \bar{\nu}_2) + \mathbb{P}_{\lambda_1,\lambda_2}(\bar{\nu}_1 < \lambda_1, \bar{\nu}_1 \geq \lambda_2, \bar{\nu}_1 \leq \bar{\nu}_2) \tag{18}$$

$$= \alpha \mathbb{P}_{\lambda_1,\lambda_2}(\nu_1 > \nu_2) + \mathbb{P}_{\lambda_1,\lambda_2}(\nu_1 < \lambda_1, \nu_1 < \lambda_2, \nu_1 \leq \nu_2) + \mathbb{P}_{\lambda_1,\lambda_2}(\lambda_2 \leq \bar{\nu}_1 < \lambda_1, \bar{\nu}_1 \leq \bar{\nu}_2) \tag{19}$$

$$\leq \alpha + \xi^{\alpha}_{\lambda_1,\lambda_2}(\bar{\nu}_1). \tag{20}$$

In lines (15) and (18) we use the following observations from the definitions of $\bar{\nu}_1$ and $\bar{\nu}_2$:

$$\{\bar{\nu}_1 > \bar{\nu}_2\} \cap \{\bar{\nu}_1 < x\} = \{\nu_1 > \nu_2\} \cap \{\max(\tilde{\nu}_1, \nu_2) < x\},$$

$$\{\bar{\nu}_1 \leq \bar{\nu}_2\} \cap \{\bar{\nu}_1 < x\} = \{\nu_1 \leq \nu_2\} \cap \{\nu_1 < x\}.$$

In line (16) we used the fact that $\tilde{\nu}_1 = \nu_S(X)$ so (a) due to this definition $\mathbb{P}_{\lambda_1,*}(\lambda_1 \leq n|\mathbf{X}^1_t) \geq 1 - \alpha$ for $t \geq \tilde{\nu}_1$ (see Eq. (1)) and (b) by conditioning on data $\mathbf{X}^1_\tau$, where $\tau = \max(\tilde{\nu}_1, \nu_1)$ the following bound holds:

$$\mathbb{P}_{\lambda_1,\lambda_2}(\tilde{\nu}_1 < \lambda_1, \nu_1 > \nu_2) = \mathbb{E}[\mathbb{P}_{\lambda_1,\lambda_2}(\tilde{\nu}_1 < \lambda_1, \nu_1 > \nu_2|\mathbf{X}^1_\tau)]$$

$$= \mathbb{E}[\mathbb{P}_{\lambda_1,\lambda_2}(\tilde{\nu}_1 < \lambda_1|\mathbf{X}^1_\tau)\mathbb{I}(\nu_1 > \nu_2)]$$

$$\leq \alpha \mathbb{P}_{\lambda_1,\lambda_2}(\nu_1 > \nu_2).$$

For line (20) a similar argument applies since (a) $\mathbb{P}_{\lambda_1,\infty}(\lambda_1 \leq n|\mathbf{X}^1_t, \mathbf{Z}^1_t) \geq 1 - \alpha$ for $t \geq \nu_1$ and (b) $\mathbb{P}_{\lambda_1,\lambda_2}(\nu_1 < \lambda_1, \nu_1 < \lambda_2, \nu_1 \leq \nu_2) = \mathbb{E}[\mathbb{P}_{\lambda_1,\infty}(\nu_1 < \lambda_1, \nu_1 < \lambda_2|\mathbf{X}^1_{\nu_2}, \mathbf{Z}^1_{\nu_2})\mathbb{I}(\nu_1 \leq \nu_2)]$.

Proceeding in a similar fashion we can obtain the result for the false alarm of subsystem 2.

**(b)** Now we can show (b) for subsystem 1. From the definition of marginal probability of false alarm in Definition 2 and following the proof steps in Eqns (15,16,18):

$$\mathbb{P}_{\lambda_1,k_2}(\bar{\nu}_1 < \lambda_1) \leq \mathbb{P}_{\lambda_1,k_2}(\tilde{\nu}_1 < \lambda_1, \nu_1 > \nu_2) + \mathbb{P}_{\lambda_1,k_2}(\bar{\nu}_1 < \lambda_1, \bar{\nu}_1 \leq \bar{\nu}_2)$$

$$\leq \alpha \mathbb{P}_{\lambda_1,k_2}(\nu_1 > \nu_2) + \mathbb{P}_{\lambda_1,k_2}(\bar{\nu}_1 < \lambda_1, \bar{\nu}_1 \leq \bar{\nu}_2).$$

The second quantity can be bound:

$$\mathbb{P}_{\lambda_1,k_2}(\bar{\nu}_1 < \lambda_1, \bar{\nu}_1 \leq \bar{\nu}_2) = \mathbb{P}_{\lambda_1,k_2}(\nu_1 < \nu_2, \nu_1 < k_1)$$

$$= \mathbb{P}_{\lambda_1,k_2}(\nu_1 < \lambda_1, \nu_1 < k_2, \nu_1 \leq \nu_2) + \mathbb{P}_{\lambda_1,k_2}(k_2 < \bar{\nu}_1 < k_1, \bar{\nu}_1 \leq \bar{\nu}_2)$$

$$\leq \alpha \mathbb{P}_{\lambda_1,k_2}(\nu_1 \leq \nu_2) + \xi_{\lambda_1,k_2}(\bar{\nu}_1).$$

## D. Proof of Theorem 3

$$\begin{aligned}
\xi^\alpha_{\lambda_1,\lambda_2}(\bar{\nu}_1) &= \sum_{k_1,k_2} \pi(k_1)\pi(k_2)\mathbb{P}_{k_1,k_2}(\bar{\nu}_1 \leq \bar{\nu}_2, k_2 \leq \bar{\nu}_1 < k_1) \\
&= \sum_{k_1,k_2} \pi(k_1)\pi(k_2)\mathbb{P}_{k_1,k_2}(\nu_1 \leq \nu_2, k_2 \leq \nu_1 < k_1) \\
&= \sum_{k_1,k_2} \pi(k_1)\pi(k_2)\mathbb{P}_{\infty,k_2}(\nu_1 \leq \nu_2, k_2 \leq \nu_1 \leq k_1) \\
&\leq \sum_{k_1,k_2} \pi(k_1)\pi(k_2)\mathbb{P}_{\infty,k_2}(k_2 \leq \nu_1 \leq \nu_2) \\
&= \sum_{k_2} \pi(k_2)\mathbb{P}_{\infty,k_2}(k_2 \leq \nu_1 \leq \nu_2).
\end{aligned}$$

We continue the proof using Lemma 1. Given this lemma, it is immediate by the dominated convergence theorem that as $\alpha \to 0$:

$$\xi^\alpha_{\lambda_1,\lambda_2}(\bar{\nu}_1) \to 0.$$

showing that the procedure is regular proving **(a)** without Assumption VII.5 . Including Assumption VII.5, **(b)** follows since $\mathbb{P}_{\infty,\lambda_2}(k_2 \leq \nu_1 \leq \nu_2) = \sum_{k_2} \pi(k_2)\mathbb{P}_{\infty,k_2}(k_2 \leq \nu_1 \leq \nu_2)$. A similar proof can be shown for subsystem 2.

## E. Lemma 1 (Event Decoupling Lemma)

**Lemma 1.** *Let Assumption VII.2. For any $k_2 > 0$, the following bound holds:*

$$\lim_{\alpha \to 0} \frac{\log \mathbb{P}_{\infty,k_2}(k_2 \leq \nu_1 \leq \nu_2)}{\log \alpha} \geq r^*.$$

*Let Assumption VII.2 and VII.5. Then:*

$$\lim_{\alpha \to 0} \frac{\log \mathbb{P}_{\infty,\lambda_2}(k_2 \leq \nu_1 \leq \nu_2)}{\log \alpha} \geq r^*.$$

*Proof:* The proof has five parts. In the first part we decompose the probability into three tail events that determine the $\alpha$-order of the error coupling probability. The point at which we switch between the first two events is a parameter ($\tilde{C}_\alpha$) that needs to be optimized. For each event we compute upper bounds to the probabilities and the rate function for the speed with which the error coupling probability converges to zero as $\alpha \to 0$. Using rate matching, we optimize the free parameter $\tilde{C}_\alpha$. Finally, we determine the parameter ($C_\alpha$), that is when one switches from the second to the third event, based on the choice of optimized parameter.

**Decomposing the event decoupling lemma into 3 events**. First notice that (we consider $C_\alpha = \infty$ a valid possibility):

$$\mathbb{P}_{\infty,k_2}(k_2 \leq \nu_1 \leq \nu_2) \leq \mathbb{P}_{\infty,k_2}(k_2 \leq \nu_1 \leq \nu_2, \nu_2 \leq k_2 + C_\alpha) + \mathbb{P}_{\infty,k_2}(\nu_2 > k_2 + C_\alpha).$$

We decompose further the quantity:

$$\mathbb{P}_{\infty,k_2}(k_2 \leq \nu_1 \leq \nu_2, \nu_2 \leq k_2 + C_\alpha) \leq \mathbb{P}_{\infty,k_2}\left(\bigcup_{l=k_2}^{k_2+C_\alpha} \{\Lambda_l(X,Z) \geq B_\alpha\} \cap \{\Lambda_l(Y,Z) < B_\alpha\}\right),$$

where the bound follows from the definition of $\nu_1$ and $\nu_2$. The advantage of this particular bound is that for small $l$, the first event - subsystem 1 mistakenly crossing the threshold -of the intersection has small probability, and for large $l$ the second does - subsystem 2 not crossing the threshold before subsystem 1. From definition of the test quantities (Definition 4), we obtain the bounds:

$$\log \Lambda_n(X,Z) \leq -\log \Pi_1(n) + \max_{r \in [1,n]} \{R_n^r(X) + R_n^r(Z)\}.$$

Now we can continue to bound:

$$\mathbb{P}_{\infty,k_2}(k_2 \leq \nu_1 \leq \nu_2, \nu_2 \leq k_2 + C_\alpha) \leq \sum_{l=k_2}^{k_2+C_\alpha} \mathbb{P}_{\infty,k_2}\left(\{\Lambda_l(X,Z) \geq B_\alpha\} \cap \{\Lambda_l(Y,Z) < B_\alpha\}\right)$$

$$\leq \sum_{l=k_2}^{k_2+C_\alpha} \mathbb{P}_{\infty,k_2}(\{-\log \Pi_1(l) + \max_{r \in [1,l]} \{R_l^r(X) + R_l^r(Z)\} \geq \log B_\alpha\} \cap$$

$$\{-\log \Pi_2(l) + \log \pi_2(r) + R_l^r(Y) + R_l^r(Z) < \log B_\alpha, \forall r \leq l\})$$

$$\leq \sum_{l=k_2}^{k_2+C_\alpha} \mathbb{P}_{\infty,k_2}(\{-\log \Pi_1(l) + \max_{r \in [1,l]} \{R_l^r(X) + R_l^r(Z)\} \geq \log B_\alpha\} \cap$$

$$\{-\log \Pi_1(l) + \log \Pi_2(l) - \log \pi_2(k_2) + \max_{r \in [1,l]} \{R_l^r(X) + R_l^r(Z)\} - R_l^{k_2}(Y) - R_l^{k_2}(Z) > \epsilon\})$$

$$\leq \sum_{l=k_2}^{k_2+\tilde{C}_\alpha} \mathbb{P}_{\infty,k_2}\left(\max_{r \in [1,l]} \{R_l^r(X) + R_l^r(Z)\} \geq \log B_\alpha + \log \Pi_1(l)\right) +$$

$$+ \sum_{l=k_2+\tilde{C}_\alpha}^{k_2+C_\alpha} \mathbb{P}_{\infty,k_2}\left(\max_{r \in [1,l]} \{R_l^r(X) + R_l^r(Z)\} - R_l^{k_2}(Y) - R_l^{k_2}(Z) > V_l\right),$$

where $V_l = \epsilon + \log \Pi_1(l) - \log \Pi_2(l) + \log \pi_2(k_2)$.

**Analyzing the probability of early crossing for subsystem 1 (event $\mathcal{E}_1$).** Lemma 2 will be used to bound the first probability in the inequality. Define $b_0 = q_0(Z) + q_0(X)$ and $b_1 = q_0(X) - q_1(Z) + d_1$. Apply Assumption VII.2 and Lemma 2, with $a = a_1 = \log B_\alpha + \log \Pi_1(l) - (l+1)d_1$, $b = b_1$, $c = c_1 = (k_2 - 1)\sigma_0^2(X,Z)$ and $d = d_1 = \sigma_1^2(X,Z)$:

$$\mathbb{P}_{\infty,k_2}\left(\max_{r \in [1,l]} \{R_l^r(X) + R_l^r(Z)\} \geq \log B_\alpha + \log \Pi_1(l)\right)$$

$$\leq l \max_{r \in [1,l]} \mathbb{P}_{\infty,k_2}(R_l^r(X) + R_l^r(Z) \geq \log B_\alpha + \log \Pi_1(l))$$

$$\leq l \max_{s \in [0,k_2-1]} \max_{r \in [k_2,l]} K(X,Z) \exp\left\{-\frac{(\log B_\alpha + \log \Pi_1(l) - (l+1)d_1 + (l-r+1)b_1 + s b_0)^2}{(l-r+1)\sigma_1^2(X,Z) + s \sigma_0^2(X,Z)}\right\}$$

$$\leq l K(X,Z) \max_{r \in [k_2,l]} \exp\left\{-\frac{(\log B_\alpha + \log \Pi_1(l) - (l+1)d_1 + (l-r+1)b_1)^2}{(l-r+1)\sigma_1^2(X,Z) + (k_2-1)\sigma_0^2(X,Z)}\right\}. \quad (21)$$

Let us assume that $\tilde{C}_\alpha = \log B_\alpha / w$ for some constant $w > 0$. We can then control the bound using

Eq. (21) and a simple observation:

$$\sum_{l=k_2}^{k_2+\tilde{C}_\alpha} \mathbb{P}_{\infty,k_2}\left(\max_{r\in[1,l]}\{R_l^r(X)+R_l^r(Z)\} \geq \log B_\alpha + \log \Pi_1(l)\right) \leq (\tilde{C}_\alpha)^2 \exp-\min_l \Phi_\alpha(l),$$

and $\Phi_\alpha$ is given by

$$\Phi_\alpha = \frac{(\log B_\alpha + A_c(K_\alpha) + K_\alpha b_1)^2}{K_\alpha \sigma_1^2(X,Z) + (k_2-1)\sigma_0^2(X,Z)},$$

$$A_c(K_\alpha) = \log \Pi_1(K_\alpha + k_2) - (K_\alpha + k_2)d_1.$$

The constant $K_\alpha$ is chosen as to minimize $\Phi_\alpha$ under the constraint that $0 < K_\alpha < \tilde{C}_\alpha$. By assumption on tail of prior, there exists $T$, such that for all $K_\alpha > T$, $|A_c(K_\alpha)| < \epsilon$. We are in this regime. Consider the case $b_1 > 0$. Our previous calculation shows minima is achieved when $K_\alpha = (\log B_\alpha - \epsilon)/b_1 - 2(k_2-1)\sigma_0^2(X,Z)/\sigma_1^2(X,Z)$. For vanishing $\alpha$, $K_\alpha < \tilde{C}_\alpha$ if $b_1 > w$, else we should set $K_\alpha = \log B_\alpha/w$ to minimize $\Phi_\alpha$. Lemma 2 can be used to compute the rate at the minimum when either $b_1 > w$ or $b_1 \leq w$:

$$\Phi_{\alpha,\min} = 4\frac{b_1^2}{\sigma_1^2(X,Z)}\left(\frac{\log B_\alpha - \epsilon}{b_1} - \frac{(k_2-1)\sigma_0^2(X,Z)}{\sigma_1^2(X,Z)}\right) \text{ for } b_1 > w,$$

$$= \frac{\left(\log B_\alpha\left[1+\frac{b_1}{w}\right] - \epsilon\right)^2}{\log B_\alpha \frac{\sigma_1^2(X,Z)}{w} + (k_2-1)\sigma_0^2(X,Z)} \text{ for } b_1 \leq w.$$

The rate that the probability goes to zero is then calculated as:

$$r_1(w) = \lim_{\alpha \to 0} \frac{-\log[(\tilde{C}_\alpha)^2 \exp-\Phi_\alpha]}{\log B_\alpha} = \begin{cases} 4\frac{b_1}{\sigma_1^2(X,Z)} & \text{for } b_1 > w, \\ \frac{w}{\sigma_1^2(X,Z)}\left[1+\frac{b_1}{w}\right]^2 & \text{for } b_1 \leq w. \end{cases} \quad (22)$$

We can proceed similarly for the case $b_1 \leq 0$. Notice that to obtain a vanishing probability now, we need $K_\alpha < \log B_\alpha/-b_1$, so the only interesting case is when $w > -b_1$ (else $\Phi_\alpha = 0$ is the minimum). Since for $b_1 < 0$, the function first decreases to the minimum, we can conclude that in this case:

$$r_1(w) = \lim_{\alpha \to 0} \frac{-\log[(\tilde{C}_\alpha)^2 \exp-\Phi_\alpha]}{\log B_\alpha} = \frac{w}{\sigma_1^2(X,Z)}\left[1+\frac{b_1}{w}\right]^2. \quad (23)$$

**Analyzing the probability of subsystem 2 crossing after subsystem 1 (event $\mathcal{E}_2$).** Let $\tilde{V}_l = \epsilon + \log \Pi_1(l) - d_1 l - \log \Pi_2(l) + d_2 l + \log \pi_2(k_2)$, $q_y(l) = (l-k_2+1)q_1(Y)$ and $\sigma_y^2(l) = (l-k_2+1)\sigma_1^2(Y)$. Similarly, for the second probability, we bound:

$$\mathbb{P}_{\infty,k_2}\left(\max_{r\in[1,l]}\{R_l^r(X)+R_l^r(Z)\} - R_l^{k_2}(Y) - R_l^{k_2}(Z) > V_l\right)$$

$$\leq l \max_{r\in[1,l]} \mathbb{P}_{\infty,k_2}\left(R_l^r(X)+R_l^r(Z) - R_l^{k_2}(Y) - R_l^{k_2}(Z) \geq V_l\right)$$

$$\leq l \max_{r\in[1,l]} \exp\left\{-\frac{(V_l + (l-r+1)q_0(X) + q_y(l) + [k_2-r]_+ q_0(Z) + [r-k_2]_+ q_1(Z))^2}{(l-r+1)\sigma_0^2(X) + \sigma_y^2(l) + [k_2-r]_+ \sigma_0^2(Z) + [r-k_2]_+ \sigma_1^2(Z)}\right\}$$

$$\leq l \max_{r\in[1,l]} \exp\left\{-\frac{(V_l + (l-r)q_0(X) + (r-k_2)q_1(Z) + q_y(l) + q_0(X))^2}{(l-r)\sigma_0^2(X) + r\sigma_1^2(Z) + \sigma_y^2(l) + k_2\sigma_0^2(Z) + \sigma_0^2(X)}\right\}$$

$$\leq l \exp\left\{-\frac{(A_e(l) + l[q_{i^*} + q_1(Y) + d_1 - d_2])^2}{C_e + l[\sigma_{i^*}^2 + \sigma_1^2(Y)]}\right\},$$

where $q_{i^*} = \min(q_0(X), q_1(Z))$, $\sigma_{i^*}^2 = \max(\sigma_0^2(X), \sigma_1^2(Z))$, $A_e(l) = \tilde{V}_l - k_2[q_1(Y) + q_1(Z)] + q_0(X) + q_1(Y)$ and $C_e = k_2[\sigma_0^2(Z) - \sigma_1^2(Y)] + \sigma_0^2(X) + \sigma_1^2(Y)$.

To continue the analysis, we compute the rates for the second major event:

$$\sum_{l=k_2+\tilde{C}_\alpha}^{k_2+C_\alpha} \mathbb{P}_{\infty,k_2}\left(\max_{r \in [1,l]}\{R_l^r(X) + R_l^r(Z)\} - R_l^{k_2}(Y) - R_l^{k_2}(Z) > V_l\right) \leq (C_\alpha - \tilde{C}_\alpha)^2 \exp-\min\tilde{\Phi}_\alpha, \text{ where}$$

$$\tilde{\Phi}_\alpha = \frac{\left(A_e(\tilde{K}_\alpha) + \tilde{K}_\alpha[q_{i^*} + q_1(Y) + d_1 - d_2]\right)^2}{C_e + \tilde{K}_\alpha[\sigma_{i^*}^2 + \sigma_1^2(Y)]}.$$

Lemma 2 implies the minimum in this case is at $\tilde{K}_\alpha = A_e(l))/(q_{i^*} + q_1(Y) + d_1 - d_2)$. But since this is a small quantity compared to $\tilde{C}_\alpha + k_2$, assuming $(q_{i^*} + q_1(Y) + d_1 - d_2) > 0$, we have that the minimum happens at $\tilde{K}_\alpha = k_2 + \tilde{C}_\alpha$, as the function is increasing after the minima. Using similar arguments as for the first major event, it is straightforward to show that the rate function satisfies:

$$r_2(w) = \lim_{\alpha \to 0} \frac{-\log[(C_\alpha - \tilde{C}_\alpha)^2 \exp-\tilde{\Phi}_\alpha]}{\log B_\alpha} = \frac{1}{w}\frac{[q_{i^*} + q_1(Y) + d_1 - d_2]^2}{\sigma_{i^*}^2 + \sigma_1^2(Y)}. \tag{24}$$

**Selecting the optimizing rate**. Given the bounds we have computed, the problem reduces to selecting the constant $\tilde{C}_\alpha$ so that the best rate is obtained for $\mathbb{P}_{\infty,k_2}(k_2 \leq \nu_1 \leq \nu_2)$. In rate matching, we have two rates $r_1(w)$ and $r_2(w)$, and would like to maximize the *minimum* of both, i.e., $\max\min(r_1(w), r_2(w))$, which is obtained by setting $w$ such that $r_1(w) = r_2(w)$, where the rate functions are given by Eq. (22), Eq. (23) (here we denote it $r_1(w)$) and Eq. (24). There are three cases, since the first event has three behaviors for the rate $r^*$:

(1) Consider $b_1 > 0$. Then, for $w < b_1$, in order to have

$$r_2(w) > 4\frac{b_1}{\sigma_1^2(X,Z)},$$

we set:

$$w_1^* < \min\left(b_1, \frac{\sigma_1^2(X,Z)}{\sigma_{i^*}^2 + \sigma_1^2(Y)}\frac{[q_{i^*} + q_1(Y) + d_1 - d_2]^2}{4b_1}\right)$$

and get rate $r^* = 4\frac{b_1}{\sigma_1^2(X,Z)}$.

(2) Again let $b_1 > 0$. Then for $w \geq b_1$, in order to have

$$r_2(w) = \frac{w}{\sigma_1^2(X,Z)}\left[1 + \frac{b_1}{w}\right]^2,$$

we set:

$$w_2^* = \sqrt{\frac{\sigma_1^2(X,Z)}{\sigma_{i^*}^2 + \sigma_1^2(Y)}}[q_{i^*} + q_1(Y) + d_1 - d_2] - b_1,$$

as long as it satisfies $w_2^* \geq b_1$. The obtained rate is $r^* = r_2(w_2^*)$. Else, set $w_2^* = b_1$, and obtain rate $r_2(b_1)$.

(3) Let $b_1 \leq 0$. Then for $w \geq -b_1$, in order to have

$$r_2(w) = \frac{w}{\sigma_1^2(X,Z)}\left[1 + \frac{b_1}{w}\right]^2,$$

we set:
$$w_3^* = \sqrt{\frac{\sigma_1^2(X,Z)}{\sigma_{i^*}^2 + \sigma_1^2(Y)}}[q_{i^*} + q_1(Y) + d_1 - d_2] - b_1,$$

which satisfies $w_3^* \geq -b_1$. The obtained rate is $r^* = r_2(w_3^*)$.

**Upper bounding detection of subsystem 2 and selecting $C_\alpha$ (probability of event $\mathcal{E}_3$).** We bound $\mathbb{P}_{\infty,k_2}(\nu_2 > k_2 + C_\alpha)$. Let Assumption VII.5 and $C_\alpha = \beta \log B_\alpha$. From definition of the test quantity (Definition 4):
$$\log \Lambda_n(Y,Z) \geq -\log \Pi_2(n) + \log \pi_2(r) + R_n^r(Y) + R_n^r(Z).$$

Let $\eta = \min\{n : R_{k_2+n-1}^{k_2}(Y) + R_{k_2+n-1}^{k_2}(Z) \geq \log B_\alpha\}$, so $\nu_2 \leq \eta$. For arbitrary $\epsilon > 0$, let
$$T_\epsilon^{k_2} = \sup\{n : |(n)^{-1}[R_{k_2+n-1}^{k_2}(Y) + R_{k_2+n-1}^{k_2}(Z)] - (q_1(Y) + q_1(Z) + d_2)| \geq \epsilon\}.$$

It is simple to see that:
$$\log B_\alpha > R_{\eta-1}^{k_2}(Y) + R_{\eta-1}^{k_2}(Z) \geq (\eta - k_2)(q_1(Y) + q_1(Z) + d_2 - \epsilon) \text{ on } \{\eta - 1 \geq T_\epsilon^{k_2}\}.$$

So,
$$\nu_2 \leq \eta \leq \left(k_2 + \frac{\log B_\alpha}{q_1(Y) + q_1(Z) + d - \epsilon}\right)\mathbb{I}(\eta < 1 + T_\epsilon^{k_2}) + (1 + T_\epsilon^{k_2})\mathbb{I}(\eta \geq 1 + T_\epsilon^{k_2})$$
$$\leq k_2 + \frac{\log B_\alpha}{q_1(Y) + q_1(Z) + d - \epsilon} + 1 + T_\epsilon^{k_2}.$$

Using this result:
$$\mathbb{P}_{\infty,k_2}(\nu_2 > k_2 + C_\alpha) \leq \mathbb{P}_{\infty,k_2}(C_\alpha \leq \frac{\log B_\alpha}{q_1(Y) + q_1(Z) + d - \epsilon} + 1 + T_\epsilon^{k_2})$$
$$\leq \mathbb{P}_{\infty,k_2}\left[T_\epsilon^{k_2} + 1 \geq \log B_\alpha\left(\beta - \frac{1}{q_1(Y) + q_1(Z) + d - \epsilon}\right)\right]$$
$$\leq \mathbb{E}_{\infty,k_2} \exp(T_\epsilon^{k_2} + 1)\left(\frac{\alpha}{1-\alpha}\right)^{\beta - \frac{1}{q_1(Y)+q_1(Z)+d_2-\epsilon}}$$
$$\leq O\left(\alpha^{\beta - \frac{1}{q_1(Y)+q_1(Z)+d_2-\epsilon}}\right).$$

where we used Markov's inequality in the last line. Assumption VII.5 guarantees that $\mathbb{E}_{\infty,k_2} \exp(T_\epsilon^{k_2} + 1) < \infty$. The constants in big-O are independent of $k_2, \epsilon$. To obtain the best possible rate for the total error coupling probability, we choose
$$\beta = (1+\epsilon)r^* + \frac{1}{q_1(Y) + q_1(Z) + d_2 - \epsilon}$$

**Concluding the proof.** To put the elements of the proof together, we use the bound:
$$\mathbb{P}_{\infty,k_2}(k_2 \leq \nu_1 \leq \nu_2) \leq \mathbb{P}_{\infty,k_2}(\mathcal{E}_1) + \mathbb{P}_{\infty,k_2}(\mathcal{E}_2) + \mathbb{P}_{\infty,k_2}(\mathcal{E}_3),$$

so the rate function has

$$\lim_{\alpha \to 0} \frac{-\log \mathbb{P}_{\infty,k_2}(k_2 \leq \nu_1 \leq \nu_2)}{\log B_\alpha} \leq \lim_{\alpha \to 0} \frac{-\log 3 \max_i \mathbb{P}_{\infty,k_2}(\mathcal{E}_i)}{\log B_\alpha}$$

$$= \min_i \lim_{\alpha \to 0} \frac{-\log \mathbb{P}_{\infty,k_2}(\mathcal{E}_i)}{\log B_\alpha} = r^*.$$

Taking the expectation with respect to $\lambda_2$, we can conclude that the results hold for the measure $\mathbb{P}_{\infty,\lambda_2}$, since $k_2$ only appears in either the denominator of the bound rates, or as $l - k_2$, but for $l > k_2$. ∎

### F. Proof of Theorem 4

We prove the first assertion. First notice that from Definition 7, $\tilde{\Delta}_1(\alpha, k_2) \subseteq \Delta_1(\alpha)$. Also notice that $\tilde{\Delta}_1(\alpha, k_2) \subseteq \tilde{\Delta}_1(\alpha)$, so that $\tilde{\Delta}_1(\alpha, k_2) \subseteq \Delta_1(\alpha) \cap \tilde{\Delta}_1(\alpha)$. Let $\nu_1 \in \tilde{\Delta}_1(\alpha, k_2)$, if $k_1 \leq k_2$:

$$\mathbb{E}_{k_1,k_2}[(\nu_1 - k_1)^m | \nu_1 \geq k_1] = \frac{\mathbb{E}_{k_1,k_2}[(\nu_1 - k_1)_+^m]}{\mathbb{P}_{k_1,k_2}(\nu_1 \geq k_1)}$$

$$\geq \frac{((1-\epsilon)L_\alpha^1)^m}{\mathbb{P}_{k_1,k_2}(\nu_1 \geq k_1)}(\mathbb{P}_{k_1,k_2}(\nu_1 \geq k_1) - \gamma_{k_1,k_2}(\nu_1)).$$

But $\mathbb{P}_{k_1,k_2}(\nu_1 \geq k_1) = 1 - \mathbb{P}_{\infty,\infty}(\nu_1 < k_1) \geq 1 - \alpha/\Pi_{k_1}^1$ for $k_1 \leq k_2$ using Lemma 3(i), and Lemma 4(ii) shows that $\gamma_{k_1,k_2}(\nu_1) \to 0$ uniformly over $\nu_1$, so

$$\inf_{\nu_1 \in \tilde{\Delta}_1(\alpha,k_2)} \mathbb{E}_{k_1,k_2}[(\nu_1 - k_1)^m | \nu_1 \geq k_1] \geq ((1-\epsilon)L_\alpha^1)^m(1 + o(1)) \quad \text{as } \alpha \to 0.$$

A similar bound works for $k_2 < k_1$, except $\mathbb{P}_{k_1,k_2}(\nu_1 \geq k_1) = \mathbb{P}_{\infty,k_2}(\nu_1 \geq k_1) \geq 1 - \alpha/\Pi_n^1$ for $\nu_1 \in \tilde{\Delta}_1(\alpha, k_2)$ (Lemma 3(ii)).

For the second statement, we note that:

$$\inf_{\nu_1 \in \tilde{\Delta}_1(\alpha)} \mathbb{E}_{\lambda_1,\lambda_2}[(\nu_1 - \lambda_1)_+^m] \geq \inf_{\nu_1 \in \tilde{\Delta}_1(\alpha)} \mathbb{E}_{\lambda_1,\lambda_2}[(\nu_1 - \lambda_1)_+^m \mathbb{I}(\lambda_1 \leq \lambda_2)]$$

$$+ \inf_{\nu_1 \in \tilde{\Delta}_1(\alpha)} \mathbb{E}_{\lambda_1,\lambda_2}[(\nu_1 - \lambda_1)_+^m \mathbb{I}(\lambda_1 > \lambda_2)]$$

We can use Lemma 4 (i) and (iii) to bound such quantities in the same manner as in the first case. Lemma 3 (i) and (iii) can be used to bound the appropriate probabilities as before.

### G. Proof of Theorem 5

We divide the proof into computing an upper bound (item (a)) and the lower bound (item (b)). First, we compute the upper bound in Lemma 5. Denote by $\nu_1 = \nu_S(X, Z)$ the stopping time given by Eqn (9). We would like to bound the expectation $\mathbb{E}_{\lambda_1,\lambda_2}[(\nu_1 - \lambda_1)^+]$. In order to do this we need Assumption VII.4. Assumption VII.4 is stronger than Assumption VII.3, and in fact the later follows from the former [25]. We can proceed to prove the theorem.

**(a)** Define:

$$q_1^d = q_1(X) + q_1(Z) + d_1, \tilde{q}_1^d = q_1(X) + d_1,$$

$$\delta_\alpha(k_1, k_2) = \mathbb{P}_{k_1,k_2}(\nu_1 > \nu_2), \mu_\alpha(k_1, k_2) = \mathbb{P}_{k_1,k_2}(\nu_1 > \tilde{\nu}_1),$$

$$\delta_\alpha = \mathbb{P}_{\lambda_1,\lambda_2}(\nu_1 > \nu_2), \mu_\alpha = \mathbb{P}_{\lambda_1,\lambda_2}(\nu_1 > \tilde{\nu}_1).$$

We start by analyzing the expectation of the stopping time, using the definition of $\bar{\nu}_1$ and $\bar{\nu}_2$:

$$\begin{aligned}
\mathbb{E}_{k_1,k_2}\left[(\bar{\nu}_1 - \lambda_1)^+\right] &= \mathbb{E}_{k_1,k_2}\left[(\nu_1 - \lambda_1)^+ \mathbb{I}(\nu_1 \leq \nu_2)\right] + \\
&+ \mathbb{E}_{k_1,k_2}\left[(\tilde{\nu}_1 - \lambda_1)^+ \mathbb{I}(\nu_1 > \nu_2, \tilde{\nu}_1 \geq \nu_2)\right] + \\
&+ \mathbb{E}_{k_1,k_2}\left[(\nu_2 - \lambda_1)^+ \mathbb{I}(\nu_1 > \nu_2 > \tilde{\nu}_1)\right]
\end{aligned}$$

Each expectation can be bounded individually. The first expectation is bounded by using Lemma 5, setting $\mathcal{A} = \{\omega \in \Omega : \nu_1(\omega) \leq \nu_2(\omega)\}$:

$$\begin{aligned}
\mathbb{E}_{k_1,k_2}[(\nu_1 - \lambda_1)^+ \mathbb{I}(\nu_1 \in \mathcal{A})]^m &\leq \left[\frac{|\log(\alpha)|}{q_1^d}\right]^m \mathbb{P}_{k_1,k_2}(\nu_1 \leq \nu_2)(1 + o(1)) \\
&= \left[\frac{|\log(\alpha)|}{q_1^d}\right]^m (1 - \delta_\alpha(k_1, k_2))(1 + o(1)).
\end{aligned}$$

For the remainder of the proof, let $\mathbb{E}$ denote $\mathbb{E}_{k_1,k_2}$ and $\mathbb{P}$ denote $\mathbb{P}_{k_1,k_2}$. We return to the usual notation wherever necessary. Also, we show the results for the case $m = 1$ and the modifications for the case $m \leq r$ are straightforward. The second expectation is bounded as:

$$\begin{aligned}
\mathbb{E}\left[(\tilde{\nu}_1 - \lambda_1)^+ \mathbb{I}(\nu_1 > \nu_2, \tilde{\nu}_1 \geq \nu_2)\right] &\leq \mathbb{E}\left[(\tilde{\nu}_1 - \lambda_1)^+ \mathbb{I}(\nu_1 > \nu_2)\right] \\
&= \mathbb{E}[(\tilde{\nu}_1 - \lambda_1)^+] - \mathbb{E}[(\tilde{\nu}_1 - \lambda_1)^+ \mathbb{I}(\nu_1 \leq \nu_2)] \\
&\leq \mathbb{E}[(\tilde{\nu}_1 - \lambda_1)^+] - \frac{\log B_\alpha}{\tilde{q}_1^d} \mathbb{P}(\tilde{\nu}_1 > \lambda_1 + \frac{\log B_\alpha}{\tilde{q}_1^d}, \nu_1 \leq \nu_2) \\
&\leq \mathbb{E}[(\tilde{\nu}_1 - \lambda_1)^+] - \frac{\log B_\alpha}{\tilde{q}_1^d} \left[\mathbb{P}(\tilde{\nu}_1 > \lambda_1 + \frac{\log B_\alpha}{\tilde{q}_1^d}) - \mathbb{P}(\nu_1 > \nu_2)\right] \\
&= \mathbb{E}[(\tilde{\nu}_1 - \lambda_1)^+] - \frac{\log B_\alpha}{\tilde{q}_1^d} \left[\mathbb{P}_{k_1,\infty}(\tilde{\nu}_1 > \lambda_1 + \frac{\log B_\alpha}{\tilde{q}_1^d}) - \mathbb{P}(\nu_1 > \nu_2)\right] \\
&\leq \mathbb{E}[(\tilde{\nu}_1 - \lambda_1)^+] - \frac{\log B_\alpha}{\tilde{q}_1^d}(1 - \tilde{\epsilon}_\alpha - \delta_\alpha(k_1, k_2)) \\
&= \mathbb{E}_{k_1,\infty}[(\tilde{\nu}_1 - \lambda_1)^+] - \frac{\log B_\alpha}{\tilde{q}_1^d}(1 - \tilde{\epsilon}_\alpha - \delta_\alpha(k_1, k_2)) \\
&\leq \frac{\log B_\alpha}{\tilde{q}_1^d}(\tilde{\epsilon}_\alpha + \delta_\alpha(k_1, k_2)).
\end{aligned}$$

Since (1) In third line we used $P(A \cap B) \geq P(A) - P(B^c)$; (2) In fifth line, $\tilde{\nu}_1$ does not depend on $k_2$; (3) $\mathbb{P}_{k_1,\infty}(\tilde{\nu}_1 > \lambda_1 + \frac{\log B_\alpha}{\tilde{q}_1^d}) \geq 1 - \tilde{\epsilon}_\alpha$, by Lemma 4 (iv) and (4) $\mathbb{E}_{k_1,\infty}[(\tilde{\nu}_1 - \lambda_1)^+]$ is bounded by Lemma 5 (fifth statement).

Finally,

$$\begin{aligned}
\mathbb{E}\left[(\nu_2 - \lambda_1)^+ \mathbb{I}(\nu_1 > \nu_2 > \tilde{\nu}_1)\right] &\leq \mathbb{E}[(\nu_1 - \lambda_1)^+ \mathbb{I}(\nu_1 > \tilde{\nu}_1)] \\
&\leq \frac{\log B_\alpha}{q_1^d} \mathbb{P}(\nu_1 > \tilde{\nu}_1)(1 + o(1)) \\
&= \frac{\log B_\alpha}{q_1^d} \mu_\alpha(k_1, k_2)(1 + o(1)).
\end{aligned}$$

Where we used (1) $\nu_1 > \nu_2$ in the first line and (2) in the second line, Lemma 5 (third assertion), setting $\mathcal{A} = \{\omega \in \Omega : \nu_1(\omega) \leq \tilde{\nu}_1(\omega)\}$.

In sum, we have:

$$\mathbb{E}[(\bar{\nu}_1 - \lambda_1)^+] \leq \frac{\log B_\alpha}{q_1^d}(1 - \delta_\alpha(k_1, k_2) + \mu_\alpha(k_1, k_2))(1 + o(1)) + \frac{\log B_\alpha}{\tilde{q}_1^d}(\tilde{\epsilon}_\alpha + \delta_\alpha(k_1, k_2))$$

$$= \frac{\log B_\alpha}{q_1^d}(1 - \delta_\alpha(k_1, k_2) + \mu_\alpha(k_1, k_2) + o(1)) + \frac{\log B_\alpha}{\tilde{q}_1^d}(\tilde{\epsilon}_\alpha + \delta_\alpha(k_1, k_2)).$$

To obtain the delay, divide

$$\mathbb{E}_{\lambda_1,\lambda_2}[(\bar{\nu}_1 - \lambda_1)^+] \leq \frac{\log B_\alpha}{q_1^d}(1 - \delta_\alpha + \mu_\alpha + o(1)) + \frac{\log B_\alpha}{\tilde{q}_1^d}(\tilde{\epsilon}_\alpha + \delta_\alpha),$$

by (using Theorem 2),

$$\mathbb{P}_{\lambda_1,\lambda_2}(\bar{\nu}_1 \geq \lambda_1) \geq 1 - \alpha - \xi^\alpha_{\lambda_1,\lambda_2}(\bar{\nu}_1)$$
$$\to 1 - o(1)$$

and we obtain the result in the Theorem since (1) $\tilde{\epsilon}_\alpha$ and $\mu_\alpha$ are $o(1)$ (Lemmas 4(iv) and 6) and (2) $\xi^\alpha_{\lambda_1,\lambda_2}(\bar{\nu}_1)$ is $o(1)$ as the procedure is regular.

We can now prove the matching lower bound for the delay.

**(b)** For the remainder of the proof, let $\mathbb{E}$ denote $\mathbb{E}_{k_1,k_2}$ and $\mathbb{P}$ denote $\mathbb{P}_{k_1,k_2}$. First notice that:

$$\mathbb{E}\left[(\bar{\nu}_1 - \lambda_1)^+\right] = \mathbb{E}\left[(\nu_1 - \lambda_1)^+\mathbb{I}(\nu_1 \leq \nu_2)\right] + \mathbb{E}\left[(\max(\tilde{\nu}_1, \nu_2) - \lambda_1)^+\mathbb{I}(\nu_1 > \nu_2)\right]$$
$$\geq \mathbb{E}\left[(\nu_1 - \lambda_1)^+\mathbb{I}(\nu_1 \leq \nu_2)\right] + \mathbb{E}\left[(\tilde{\nu}_1 - \lambda_1)^+\mathbb{I}(\nu_1 > \nu_2)\right].$$

We can now bound the first term.

$$\mathbb{E}\left[(\nu_1 - \lambda_1)^+\mathbb{I}(\nu_1 \leq \nu_2)\right] \geq L_1^\alpha \, \mathbb{P}(\nu_1 - k_1 > (1-\epsilon)L_1^\alpha, \nu_1 \leq \nu_2)$$
$$= L_1^\alpha \left[\mathbb{P}(\nu_1 \geq k_1 \wedge k_2, \nu_1 \leq \nu_2) - \mathbb{P}(k_2 < \nu_1 \leq k_1, \nu_1 \leq \nu_2)\right.$$
$$\left. - \mathbb{P}(k_1 < \nu_1 \leq k_1 + (1-\epsilon)L_1^\alpha, \nu_1 \leq \nu_2)\right]$$
$$\geq L_1^\alpha \left[\mathbb{P}(\nu_1 \geq k_1 \wedge k_2, \nu_1 \leq \nu_2) - \xi^\alpha_{k_1,k_2}(\nu_1) - \gamma^{(k_1,k_2)}_{\epsilon,\alpha}(\nu_1)\right]$$
$$\geq L_1^\alpha \left[\mathbb{P}(\nu_1 \geq k_1 \wedge k_2) - \mathbb{P}(\nu_1 > \nu_2) - \xi^\alpha_{k_1,k_2}(\nu_1) - \gamma^{(k_1,k_2)}_{\epsilon,\alpha}(\nu_1)\right]$$
$$= L_1^\alpha \left[1 - \mathbb{P}_{\infty,\infty}(\nu_1 \leq k_1 \wedge k_2) - \delta_\alpha(k_1, k_2) - \xi^\alpha_{k_1,k_2}(\nu_1) - \gamma^{(k_1,k_2)}_{\epsilon,\alpha}(\nu_1)\right]$$
$$\geq L_1^\alpha \left[1 - \frac{\alpha}{\Pi^1_{k_1 \wedge k_2}} - \delta_\alpha(k_1, k_2) - \xi^\alpha_{k_1,k_2}(\nu_1) - \gamma^{(k_1,k_2)}_{\epsilon,\alpha}(\nu_1)\right].$$

Where in the (1) fourth line we get a lower bound since the subtracted probabilities, and we identify them with previous definitions; (2) fifth line we use $P(A \cap B) \geq P(A) - P(B^c)$; (3) sixth line we use a change of measure and (4) seventh line we use Lemma 3(i).

So if the procedure is $(k_1, k_2)$ regular, $\mathbb{E}\left[(\nu_1 - \lambda_1)^+\mathbb{I}(\nu_1 \leq \nu_2)\right] \geq L_1^\alpha\left(1 - \delta_\alpha(k_1, k_2) + o(1)\right)$. For the averaged case over the priors, the last line above should be replaced using the false alarm bound for $\nu_1$. Notice that $\mathbb{P}(\nu_1 \leq k_1 \wedge k_2) \leq \mathbb{P}_{k_1,\infty}(\nu_1 \leq k_1)$, and average the last statement over $k_1$ to obtain $\mathbb{P}_{\lambda_1,\infty}(\nu_1 \leq \lambda_1) \leq \alpha$, so the last line is replaced by

$$\geq L_1^\alpha \left[1 - \alpha - \delta_\alpha - \xi^\alpha_{\lambda_1,\lambda_2}(\nu_1) - \gamma_{\epsilon,\alpha}(\nu_1)\right]$$

The second expectation can be bound similarly:

$$\begin{aligned} \mathbb{E}\left[(\tilde{\nu}_1 - \lambda_1)^+ \mathbb{I}(\nu_1 > \nu_2)\right] &\geq \tilde{L}_1^\alpha \, \mathbb{P}(\tilde{\nu}_1 - k_1 \geq \tilde{L}_1^\alpha, \nu_1 > \nu_2) \\ &\geq \tilde{L}_1^\alpha \left[\mathbb{P}(\tilde{\nu}_1 - k_1 \geq \tilde{L}_1^\alpha) - \mathbb{P}(\nu_1 \leq \nu_2)\right] \\ &= \tilde{L}_1^\alpha \left[1 - \mathbb{P}(\tilde{\nu}_1 < k_1) - \gamma_{\epsilon,\alpha}^{(k_1,k_2)}(\tilde{\nu}_1) - 1 + \delta_\alpha(k_1, k_2)\right]. \end{aligned}$$

Finally, we use the trivial upper bound $\mathbb{P}_{k_1,k_2}(\tilde{\nu}_1 \geq k_1) \leq 1 - o(1)$ and take expectation with respect to $\lambda_1$ and $\lambda_2$ to obtain the result in the theorem.

## APPENDIX

### A. Lemma 2

**Lemma 2.** *Consider the function $f(x) = (a + b\,x)^2/(c + d\,x)$, with $a, c, d \geq 0$. The following properties hold:*

(a) *If $b > 0$ and $a/b > 2\,c/d$, the function is decreasing in the interval $x \in [0, x_{min}]$ and increasing in $x \in (x_{min}, \infty]$, where $x_{min} = a/b - 2\,c/d$ is the point of minimum and $f(x_{min}) = 4\,b^2/d\,(a/b - c/d)$.*

(a') *If $b > 0$ and $a/b \leq 2\,c/d$, the function is increasing in the interval $x \in [0, \infty)$, the point of minimum is $x = 0$, and $f(x_{min}) = a^2/c$.*

(b) *If $b \leq 0$, the function is decreasing in the interval $x \in [0, x_{min}]$ and increasing in $x \in (x_{min}, \infty]$, where $x_{min} = -a/b$ is the point of minimum and $f(x_{min}) = 0$.*

*Proof:* Follows from noticing that the derivative is $f'(x) = (a + b\,x)(2\,bc - ad + bd\,x)/(c + d\,x)^2$. ∎

### B. Lemma 3

**Lemma 3.** *Let $\nu_1$ be a valid stopping time such that $\nu_1 \in \mathcal{F}_n(X, Z)$. Consider the stopping rule classes in Definition 7. Then:*

(i) *If $\nu_1 \in \Delta_1(\alpha)$, then for all $n \leq k_1 \leq k_2$ $\mathbb{P}_{k_1,k_2}(\nu_1 < n) \leq \frac{\alpha}{\Pi_n^1}$.*

(ii) *If $\nu_1 \in \tilde{\Delta}_1(\alpha, k_2)$, then for all $n \leq k_1$: $\mathbb{P}_{k_1,k_2}(\nu_1 < n) \leq \frac{\alpha}{\Pi_n^1}$.*

(iii) *If $\nu_1 \in \tilde{\Delta}_1(\alpha)$, then for all $n \leq k_1$: $\mathbb{P}_{k_1,\lambda_2}(\nu_1 < n, \lambda_2 < k_1) \leq \frac{\alpha}{\Pi_n^1}$.*

*Proof:* All assertions follow the same proof guideline. First notice that:

$$\begin{aligned}
P_{\text{fa}}^{\pi_1,\infty}(\nu_1) &\geq \mathbb{P}_{\lambda_1,\infty}(\{\nu_1 < n\} \cap \{\lambda_1 > n\}) \\
&= \mathbb{P}_{\lambda_1,\infty}(\nu_1 < n | \lambda_1 > n)\mathbb{P}_{\lambda_1,\infty}(\lambda_1 > n) \\
&= \mathbb{P}_{\infty,\infty}(\nu_1 < n)\Pi_n^1.
\end{aligned}$$

Next, as $\nu_1(X, Z) \in \Delta_1(\alpha)$, we have $P_{\text{fa}}^{\pi_1,\infty}(\nu_1) \leq \alpha$. To conclude, for the choices of $k_1, k_2$ and $n$ in the lemma $\mathbb{P}_{k_1,k_2}(\nu_1 < n) = \mathbb{P}_{\infty,\infty}(\nu_1 < n)$. ∎

### C. Lemma 4

We state a basic Lemma that is used to bound probabilities of false alarm in a given class. Compare this to Lemma 1 in [25].

**Lemma 4.** *Define for all $0 < \epsilon < 1$:*

$$\gamma_{\epsilon,\alpha}^{(k_1,k_2)}(\nu_1) = \mathbb{P}_{k_1,k_2}(k_1 \leq \nu_1 \leq k_1 + (1-\epsilon)L_1^\alpha),$$
$$\gamma_{\epsilon,\alpha}(\nu_1) = \mathbb{P}_{\lambda_1,\lambda_2}(\lambda_1 \leq \nu_1 \leq \lambda_1 + (1-\epsilon)L_1^\alpha),$$
$$\gamma_{\epsilon,\alpha}(\nu_1, \lambda_2 < \lambda_1) = \mathbb{P}_{\lambda_1,\lambda_2}(\lambda_1 \leq \nu_1 \leq \lambda_1 + (1-\epsilon)L_1^\alpha, \lambda_2 < \lambda_1),$$
$$\tilde{\gamma}_{\epsilon,\alpha}^{(k_1,k_2)}(\nu_1) = \mathbb{P}_{k_1,k_2}(k_1 \leq \nu_1 \leq k_1 + (1-\epsilon)\tilde{L}_1^\alpha),$$
$$\tilde{\gamma}_{\epsilon,\alpha}(\nu_1) = \mathbb{P}_{\lambda_1,\lambda_2}(\lambda_1 \leq \nu_1 \leq \lambda_1 + (1-\epsilon)\tilde{L}_1^\alpha, \lambda_2 < \lambda_1),$$
$$\gamma_{\epsilon,\alpha}^{(k_1,k_2)}(\tilde{\nu}_1) = \mathbb{P}_{k_1,k_2}(k_1 \leq \tilde{\nu}_1 \leq k_1 + (1-\epsilon)\tilde{L}_1^\alpha),$$
$$\gamma_{\epsilon,\alpha}(\tilde{\nu}_1) = \mathbb{P}_{\lambda_1,\lambda_2}(\lambda_1 \leq \tilde{\nu}_1 \leq \lambda_1 + (1-\epsilon)\tilde{L}_1^\alpha).$$

*where $d_1$ is given in Assumption VII.1, $L_1^\alpha$ and $\tilde{L}_1^\alpha$ are given by Definition 6. Then for all $k_1, k_2 \geq 1$ and $0 < \epsilon < 1$:*

(i) $\lim_{\alpha \to 0} \sup_{\nu_1 \in \Delta_1(\alpha)} \gamma_{\epsilon,\alpha}^{(k_1,k_2)}(\nu_1) = 0$,
$\lim_{\alpha \to 0} \sup_{\nu_1 \in \Delta_1(\alpha)} \gamma_{\epsilon,\alpha}(\nu_1) = 0$,
$\lim_{\alpha \to 0} \sup_{\nu_1 \in \Delta_1(\alpha)} \gamma_{\epsilon,\alpha}(\nu_1, \lambda_1 < \lambda_2) = 0$,

(ii) $\lim_{\alpha \to 0} \sup_{\nu_1 \in \tilde{\Delta}_1(\alpha,k_2)} \tilde{\gamma}_{\epsilon,\alpha}^{(k_1,k_2)}(\nu_1) = 0$ for $k_1 > k_2$,
$\lim_{\alpha \to 0} \sup_{\nu_1 \in \tilde{\Delta}_1(\alpha,k_2)} \gamma_{\epsilon,\alpha}^{(k_1,k_2)}(\nu_1) = 0$ for $k_1 \leq k_2$,

(iii) $\lim_{\alpha \to 0} \sup_{\nu_1 \in \tilde{\Delta}_1(\alpha)} \tilde{\gamma}_{\epsilon,\alpha}(\nu_1) = 0$,

(iv) $\lim_{\alpha \to 0} \sup_{\tilde{\nu}_1 \in \Delta_1(\alpha)} \gamma_{\epsilon,\alpha}^{(k_1,k_2)}(\tilde{\nu}_1) = 0$,
$\lim_{\alpha \to 0} \sup_{\tilde{\nu}_1 \in \Delta_1(\alpha)} \gamma_{\epsilon,\alpha}(\tilde{\nu}_1) = 0$.

*An analogous result holds for $\nu_2$ belonging to classes $\Delta_2(\alpha)$, $\tilde{\Delta}_2(\alpha, k_2)$ and $\tilde{\Delta}_2(\alpha)$.*

*Proof: (i)* We can first build our bound by a change of measure argument:

$$\mathbb{P}_{\infty,\infty}\left(k_1 \leq \nu_1 < k_1 + (1-\epsilon)L_1^\alpha\right) =$$
$$= \mathbb{E}_{k_1,k_2}\left\{\mathbb{I}\left(k_1 \leq \nu < k_1 + (1-\epsilon)L_1^\alpha\right) e^{-\left(R_{\nu_1}^{k_1}(X) + R_{\nu_1}^{k_1 \wedge k_2}(Z)\right)}\right\}$$
$$\geq \mathbb{E}_{k_1,k_2}\left\{\mathbb{I}_{\left(k_1 \leq \nu < k_1 + (1-\epsilon)L_1^\alpha, R_{\nu_1}^{k_1}(X) + R_{\nu_1}^{k_1 \wedge k_2}(Z) < C\right)} e^{-\left(R_{\nu_1}^{k_1}(X) + R_{\nu_1}^{k_1 \wedge k_2}(Z)\right)}\right\}$$
$$\geq e^{-C} \mathbb{P}_{k_1,k_2}\left(k_1 \leq \nu < k_1 + (1-\epsilon)L_1^\alpha, \max_{k_1 \leq n < k_1 + (1-\epsilon)L_1^\alpha} R_n^{k_1}(X) + R_n^{k_1 \wedge k_2}(Z) < C\right)$$
$$\geq e^{-C}\left[\mathbb{P}_{k_1,k_2}\left(k_1 \leq \nu < k_1 + (1-\epsilon)L_1^\alpha\right) - \right.$$
$$\left. \mathbb{P}_{k_1,k_2}\left(\max_{k_1 \leq n < k_1 + (1-\epsilon)L_1^\alpha} R_n^{k_1}(X) + R_n^{k_1 \wedge k_2}(Z) \geq C\right)\right].$$

Choosing $C = (1-\epsilon^2)(q_1(X) + q_1(Z))L_1^\alpha$, and rearranging we obtain:

$$\gamma_{\epsilon,\alpha}^{(k_1,k_2)} \leq e^{(1-\epsilon^2)(q_1(X)+q_1(Z))L_1^\alpha} \mathbb{P}_{\infty,\infty}\left(k_1 \leq \nu_1 < k_1 + (1-\epsilon)L_1^\alpha\right) + \quad (25)$$
$$+ \mathbb{P}_{k_1,k_2}\left(\max_{k_1 \leq n < k_1 + (1-\epsilon)L_1^\alpha} R_n^{k_1}(X) + R_n^{k_1 \wedge k_2}(Z) \geq C\right)$$

We now analyze each of the two parts in the above. We start with the second term:

$$\beta_{k_1,k_2}(\epsilon,\alpha) = \mathbb{P}_{k_1,k_2}\left(\max_{k_1 \leq n < k_1 + (1-\epsilon)L_1^\alpha} R_n^{k_1}(X) + R_n^{k_1 \wedge k_2}(Z) \geq C\right)$$

$$
\begin{aligned}
&\leq \mathbb{P}_{k_1,k_2}\left(\max_{k_1\leq n<k_1+(1-\epsilon)L_1^\alpha} R_n^{k_1}(X) + R_n^{k_1}(Z) \geq C\right) + \\
&+ \mathbb{P}_{k_1,k_2}\left(C - R_Z \leq \max_{k_1\leq n<k_1+(1-\epsilon)L_1^\alpha} R_n^{k_1}(X) + R_n^{k_1}(Z) < C, R_Z \geq 0\right) \\
&\leq \mathbb{P}_{k_1,k_2}\left(\max_{k_1\leq n<k_1+(1-\epsilon)L_1^\alpha} R_n^{k_1}(X) + R_n^{k_1}(Z) \geq C\right) + \\
&+ \mathbb{P}_{k_1,k_2}\left(C - R_Z \leq \max_{k_1\leq n<k_1+(1-\epsilon)L_1^\alpha} R_n^{k_1}(X) + R_n^{k_1}(Z) < C\right) \\
&= \mathbb{P}_{k_1,k_2}\left(\max_{0\leq n<(1-\epsilon)L_1^\alpha} R_{k_1+n}^{k_1}(X) + R_{k_1+n}^{k_1}(Z) \geq C\right) + \\
&+ \mathbb{P}_{k_1,k_2}\left(C - R_Z \leq \max_{0\leq n<(1-\epsilon)L_1^\alpha} R_{k_1+n}^{k_1}(X) + R_n^{k_1}(Z) < C\right) \\
&= \mathbb{P}_{k_1,k_2}\left(\frac{1}{N_\alpha}\max_{0\leq n<N_\alpha} R_{k_1+n}^{k_1}(X) + R_{k_1+n}^{k_1}(Z) \geq q_\epsilon\right) + \\
&+ \mathbb{P}_{k_1,k_2}\left(q_\epsilon - \frac{R_Z}{N_\alpha} \leq \max_{0\leq n<N_\alpha} R_{k_1+n}^{k_1}(X) + R_n^{k_1}(Z) < q_\epsilon\right).
\end{aligned}
$$

Where $R_Z = R_{k_1-1}^{k_2}(Z)$, $q_\epsilon = (1+\epsilon)(q_1(X) + q_1(Z))$ and $N_\alpha = \lfloor(1-\epsilon)L_1^\alpha\rfloor$. Now noticing that as $\alpha \to 0$ we have $N_\alpha \to \infty$, we have using assumption VII.3 and properties of measure:

$$
\begin{aligned}
\mathbb{P}_{k_1,k_2}\left(\frac{1}{N_\alpha}\max_{0\leq n<N_\alpha} R_{k_1+n}^{k_1}(X) + R_{k_1+n}^{k_1}(Z) \geq q_\epsilon\right) \\
= \mathbb{P}_{k_1,\infty}\left(\frac{1}{N_\alpha}\max_{0\leq n<N_\alpha} R_{k_1+n}^{k_1}(X) + R_{k_1+n}^{k_1}(Z) \geq q_\epsilon\right) \to 0.
\end{aligned}
$$

Because $\frac{R_Z}{N_\alpha} \to 0$ almost surely, we have the second probability going to zero. Thus $\beta_{k_1,k_2}(\epsilon, \alpha) \to 0$ as $\alpha \to 0$. We now proceed to bound the first probability in Eq. (25), using the result from Lemma 3(i) and using the definition of $N_\alpha$ and $q = q_1(X) + q_1(Z)$:

$$
\begin{aligned}
p_{k_1,k_2}(\epsilon, \alpha) &= e^{(1-\epsilon^2)(q_1(X)+q_1(Z))L_1^\alpha}\mathbb{P}_{\infty,\infty}\left(k_1 \leq \nu_1 < k_1 + (1-\epsilon)L_1^\alpha\right) \\
&\leq e^{(1-\epsilon^2)(q_1(X)+q_1(Z))L_1^\alpha}\mathbb{P}_{\infty,\infty}\left(\nu_1 < k_1 + (1-\epsilon)L_1^\alpha\right) \\
&\leq \frac{\alpha}{\Pi^1_{k_1+N_\alpha}}e^{(1-\epsilon^2)qL_1^\alpha}.
\end{aligned}
$$

Notice that $\alpha = e^{-(q+d_1)L_1^\alpha}$ from the definitions. Thus:

$$
\begin{aligned}
\frac{\log(p_{k_1,k_2}(\epsilon, \alpha))}{N_\alpha} &\leq \frac{(1-\epsilon^2)qL_1^\alpha}{N_\alpha} - \frac{(q+d_1)L_1^\alpha}{N_\alpha} - \frac{\log \Pi^1_{k_1+N_\alpha}}{N_\alpha} \\
&= \frac{(1-\epsilon^2)qL_1^\alpha}{N_\alpha} - \frac{(q+d_1)L_1^\alpha}{N_\alpha} - \frac{\log \Pi^1_{k_1+N_\alpha}}{k_1+N_\alpha}\frac{k_1+N_\alpha}{N_\alpha} \\
&\leq \frac{(1+\epsilon)q(N_\alpha+1)}{N_\alpha} - \frac{\frac{(q+d_1)}{1-\epsilon}N_\alpha}{N_\alpha} - \frac{\log \Pi^1_{k_1+N_\alpha}}{k_1+N_\alpha}\frac{k_1+N_\alpha}{N_\alpha} \\
&= -\frac{\epsilon^2 q + d_1}{1-\epsilon} - \frac{\log \Pi^1_{k_1+N_\alpha}}{k_1+N_\alpha}\left(1+\frac{k_1}{N_\alpha}\right) + \frac{(1+\epsilon)q}{N_\alpha}.
\end{aligned}
$$

Taking limits, and using the tail assumption:

$$\lim_{\alpha \to 0} \frac{\log(p_{k_1,k_2}(\epsilon,\alpha))}{N_\alpha} \leq -\frac{\epsilon^2 q + d_1}{1-\epsilon} + d_1 = -\frac{\epsilon^2 q + \epsilon d_1}{1-\epsilon}.$$

It is now clear that $p_{k_1,k_2}(\epsilon,\alpha) \to 0$. We have shown that for all $\nu_1 \in \Delta_1(\alpha)$:

$$\gamma_{\epsilon,\alpha}^{(k_1,k_2)}(\nu_1) \leq \beta_{k_1,k_2}(\epsilon,\alpha) + p_{k_1,k_2}(\epsilon,\alpha). \to 0$$

We can complete the result by studying the behavior of $\gamma_{\epsilon,\alpha}$. Let $\tilde{N}_\alpha = \lfloor \epsilon L_1^\alpha \rfloor$. From the definition:

$$\begin{aligned}
\gamma_{\epsilon,\alpha}(\nu_1) &= \sum_{k_1=1}^{\infty} \sum_{k_2=1}^{\infty} \pi_1(k_1)\pi_2(k_2)\gamma_{\epsilon,\alpha}^{(k_1,k_2)}(\nu_1) \\
&\leq \Pi^1_{\tilde{N}_\alpha} + \sum_{k_1=1}^{\tilde{N}_\alpha} \sum_{k_2=1}^{\tilde{N}_\alpha} \pi_1(k_1)\pi_2(k_2)(\beta_{k_1,k_2}(\epsilon,\alpha) + p_{k_1,k_2}(\epsilon,\alpha)) \\
&\leq \Pi^1_{\tilde{N}_\alpha} + \sup_{k_1 \leq \tilde{N}_\alpha} p_{k_1,k_2}(\epsilon,\alpha) + \sum_{k_1=1}^{\tilde{N}_\alpha} \sum_{k_2=1}^{\tilde{N}_\alpha} \pi_1(k_1)\pi_2(k_2)\beta_{k_1,k_2}(\epsilon,\alpha).
\end{aligned}$$

Now as $\alpha \to 0$, $\Pi^1_{\tilde{N}_\alpha} \to 0$ by definition, and the third term in the above sum goes to zero by Dominated Convergence Theorem and the fact that $\beta_{k_1,k_2}(\epsilon,\alpha) \to 0$. For the second term, we make a minor modification in the first proof of convergence of $p_{k_1,k_2}(\epsilon,\alpha)$, by noticing that $\Pi^1_n$ is a non-increasing function of $n$:

$$\sup_{k_1 \leq \tilde{N}_\alpha} p_{k_1,k_2}(\epsilon,\alpha) \leq \frac{\alpha}{\Pi^1_{\tilde{N}_\alpha + N_\alpha}} e^{(1-\epsilon^2)qL_1^\alpha}.$$

Then continuing as before, replacing $k_1$ by $\tilde{N}_\alpha$, we obtain:

$$\lim_{\alpha \to 0} \frac{\log(\sup_{k_1 \leq \tilde{N}_\alpha} p_{k_1,k_2}(\epsilon,\alpha))}{N_\alpha} \leq -\frac{\epsilon^2 q + d_1}{1-\epsilon} + d_1\left(1 + \frac{\epsilon}{1-\epsilon}\right) = -\frac{\epsilon^2 q}{1-\epsilon}.$$

Clearly this shows that $\sup_{k_1 \leq \tilde{N}_\alpha} p_{k_1,k_2}(\epsilon,\alpha) \to 0$, concluding the proof. The proof for the third statement is the same the above, except the sum over the priors is only over the cases $\lambda_1 < \lambda_2$.

*(ii)* The proof is as in (i), except we use the change of measure for $k_2 < k_1$:

$$\mathbb{P}_{\infty,k_2}\left(k_1 \leq \nu_1 < k_1 + (1-\epsilon)\tilde{L}_1^\alpha\right) = \mathbb{E}_{k_1,k_2}\left\{\mathbb{I}\left(k_1 \leq \nu < k_1 + (1-\epsilon)\tilde{L}_1^\alpha\right)e^{-\left(R_{\nu_1}^{k_1}(X)\right)}\right\}.$$

For $k_1 \leq k_2$ we use the same change of measure as in (i). We again can use Lemma 3(ii). For the cases (iii) and (iv) the proofs proceed similarly. ∎

## D. Lemma 5

**Lemma 5.** *Let the stopping time $\nu_1 = \nu(X, Z)$ defined in Eq. (9). If Assumption VII.4, then as $\alpha \to 0$, for all $m \leq r$, and all events $\mathcal{A}$:*

$$\mathbb{E}_{k_1,k_2}[(\nu_1 - \lambda_1)^+]^m \leq \left[L_\alpha^1\right]^m (1 + o(1)),$$

$$\mathbb{E}_{\lambda_1,\lambda_2}[(\nu_1 - \lambda_1)^+]^m \leq \left[L_\alpha^1\right]^m (1 + o(1)),$$

$$\mathbb{E}_{k_1,k_2}[(\nu_1 - \lambda_1)^+ \mathbb{I}(\nu_1 \in \mathcal{A})]^m \leq \left[L_\alpha^1\right]^m \mathbb{P}_{k_1,k_2}(\nu_1 \in \mathcal{A})(1 + o(1)),$$

$$\mathbb{E}_{\lambda_1,\lambda_2}[(\nu_1 - \lambda_1)^+ \mathbb{I}(\nu_1 \in \mathcal{A})]^m \leq \left[L_\alpha^1\right]^m \mathbb{P}_{\lambda_1,\lambda_2}(\nu_1 \in \mathcal{A})(1 + o(1)),$$

$$\mathbb{E}_{k_1,k_2}[(\tilde{\nu}_1 - \lambda_1)^+]^m \leq \left[\tilde{L}_\alpha^1\right]^m (1 + o(1)),$$

$$\mathbb{E}_{\lambda_1,\lambda_2}[(\tilde{\nu}_1 - \lambda_1)^+]^m \leq \left[\tilde{L}_\alpha^1\right]^m (1 + o(1)),$$

$$\mathbb{E}_{k_1,k_2}[(\tilde{\nu}_1 - \lambda_1)^+ \mathbb{I}(\tilde{\nu}_1 \in \mathcal{A})]^m \leq \left[\tilde{L}_\alpha^1\right]^m \mathbb{P}_{k_1,k_2}(\tilde{\nu}_1 \in \mathcal{A})(1 + o(1)),$$

$$\mathbb{E}_{\lambda_1,\lambda_2}[(\tilde{\nu}_1 - \lambda_1)^+ \mathbb{I}(\tilde{\nu}_1 \in \mathcal{A})]^m \leq \left[\tilde{L}_\alpha^1\right]^m \mathbb{P}_{\lambda_1,\lambda_2}(\tilde{\nu}_1 \in \mathcal{A})(1 + o(1)),$$

*where $L_\alpha^1$ and $\tilde{L}_\alpha^1$ are given in Definition 6.*

*Proof:* By definition of $\nu_1$, since we are using the SRP statistic:

$$\log(\Lambda_n(X, Z)) \geq \log\left(\frac{\pi_1(k_1)}{\Pi_n^1}\right) + R_n^{k_1}(X) + R_n^{k_1}(Z)$$
$$= S_n^{k_1}.$$

We can define a stopping time:

$$\eta(k_1) = \inf\left\{n : S_{k_1+n-1}^{k_1} \geq \log(B_\alpha)\right\}.$$

Notice that $\nu_1 - k_1 \leq \eta(k_1)$ on $\nu_1 \geq k_1$, as $\eta(k_1)$ starts at $k_1$ and the Shiryaev statistics only includes values in range $(k_1, n)$ after time $k_1$. Define:

$$\tilde{T}_\epsilon^{(k_1)} = \sup\left\{n \geq 1 : \left|\frac{1}{n}S_{k_1+n-1}^{k_1}(X) - q_1(X) + q_1(Z) + d_1\right| > \epsilon\right\}.$$

Due to Assumption VII.4, and because $\frac{1}{n}\log\left(\frac{\pi_1(k_1)}{\Pi_n^1}\right) \to d_1$ as $n \to \infty$, we have $\mathbb{E}_{k_1,k_2}[\tilde{T}_\epsilon^{(k_1)}] < \infty$.
Furthermore, from the definition of $\eta$ and setting $q_d = q_1(X) + q_1(Z) + d_1$:

$$\log(B_\alpha) \geq S_{\eta(k_1)-1}^{k_1} \geq (\eta(k_1) - k_1)(q_d - \epsilon) \text{ on } \left\{\eta(k) - 1 > \tilde{T}_\epsilon^{(k_1)}\right\}.$$

We can bound for all $0 < \epsilon < q_d$:

$$\eta(k_1) \leq k_1 + \frac{\log(B_\alpha)}{q_d - \epsilon}\mathbb{I}_{\{\eta(k)-1>\tilde{T}_\epsilon^{(k_1)}\}} + (\tilde{T}_\epsilon^{(k_1)} + 1)\mathbb{I}_{\{\eta(k)-1\leq\tilde{T}_\epsilon^{(k_1)}\}}$$
$$\leq \tilde{T}_\epsilon^{(k_1)} + 1 + k_1 + \frac{\log(B_\alpha)}{q_d - \epsilon}.$$

So:

$$\frac{\nu_1 - k_1}{\log(B_\alpha)} \leq \frac{\tilde{T}_\epsilon^{(k_1)}}{\log(B_\alpha)} + \frac{1 + k_1}{\log(B_\alpha)} + \frac{1}{q_d - \epsilon}.$$

Letting $\epsilon \to 0$, noticing $\mathbb{E}_{k_1,k_2}[\tilde{T}_\epsilon^{(k_1)}] < \infty$, and letting $\alpha \to 0$ ($\log(B_\alpha) \to \infty$) we obtain the first result in the Theorem for all $m \leq r$. Averaging over the priors, noticing $\mathbb{E}_{\lambda_1,\lambda_2}[\tilde{T}_\epsilon^{(k_1)}] < \infty$ we obtain the second. For the third and fourth results, it suffices to notice that:

$$\frac{(\nu_1 - k_1)\mathbb{I}(\nu_1 \in \mathcal{A})}{\log(B_\alpha)} \leq \frac{\tilde{T}_\epsilon^{(k_1)}}{\log(B_\alpha)} + \frac{1 + k_1}{\log(B_\alpha)} + \frac{\mathbb{I}(\nu_1 \in \mathcal{A})}{q_d - \epsilon}.$$

The proof follows along similar lines for $\tilde{\nu}_1$. ∎

*E. Lemma 6*

**Lemma 6.** *Let $\mu_\alpha(k_1, k_2) = \mathbb{P}_{k_1,k_2}(\nu_1 > \tilde{\nu}_1)$ and $\mu_\alpha = \mathbb{P}_{\lambda_1,\lambda_2}(\nu_1 > \tilde{\nu}_1)$, where $\nu_1$ and $\tilde{\nu}_1$ are given in Eqns. (9) and (10). Then $\mu_\alpha(k_1, k_2) = o(1)$ and $\mu_\alpha = o(1)$ as $\alpha \to 0$.*

*Proof:* First, we note that

$$\mathbb{P}_{k_1,k_2}(\nu_1 > \tilde{\nu}_1) \leq \mathbb{P}_{k_1,k_2}(\nu_1 > \tilde{\nu}_1, \tilde{\nu}_1 \geq k_1 + \tilde{L}_\alpha) + \mathbb{P}_{k_1,k_2}(\tilde{\nu}_1 < k_1) + \mathbb{P}_{k_1,k_2}(k_1 \leq \tilde{\nu}_1 \leq k_1 + \tilde{L}_\alpha),$$

and asymptotically, in $\alpha$, the last two terms are $o(1)$. Next, we follow along the lines of the first part of Lemma 1, to derive the result. Let $\mathbb{P}$ denote $\mathbb{P}_{k_1,k_2}$, $\mathcal{E}(X) = \{\log \Lambda_l(X) \geq \log B_\alpha\}$ and $\mathcal{I}(\tilde{L}_\alpha) = [k_1 + \tilde{L}_\alpha, \infty)$:

$$\mathbb{P}_{k_1,k_2}(\nu_1 > \tilde{\nu}_1, \tilde{\nu}_1 \geq k_1 + \tilde{L}_\alpha) \leq \sum_{l=k_1+\tilde{L}_\alpha}^{\infty} \mathbb{P}\left(\{\log \Lambda_l(X, Z) \leq \log B_\alpha\} \cap \mathcal{E}(X), \tilde{\nu}_1 = l\right)$$

$$\leq \sum_{l=k_1+\tilde{L}_\alpha}^{\infty} \mathbb{P}\left(\left\{\log \Lambda_l(X) + \min_{s \in [1,l]} R_l^s(Z) - \log \Pi_1(l) \leq \log B_\alpha\right\} \cap \mathcal{E}(X), \tilde{\nu}_1 = l\right)$$

$$\leq \sum_{l=k_1+\tilde{L}_\alpha}^{\infty} \mathbb{P}\left(\max_{s \in [1,l]} -R_l^s(Z) \geq -\log \Pi_1(l), \tilde{\nu}_1 = l\right)$$

$$\leq \sum_{l=k_1+\tilde{L}_\alpha}^{\infty} \mathbb{P}(\tilde{\nu}_1 = l)\mathbb{P}\left(\max_{s \in [1,l]} -R_l^s(Z) \geq -\log \Pi_1(l)\right)$$

$$\leq \max_{l \in \mathcal{I}(\tilde{L}_\alpha)} \mathbb{P}\left(\max_{s \in [1,l]} -R_l^s(Z) \geq -\log \Pi_1(l)\right)$$

$$\leq \max_{l \in \mathcal{I}(\tilde{L}_\alpha)} l \max_{s \in [1,l]} \mathbb{P}(-R_l^s(Z) \geq -\log \Pi_1(l))$$

$$\leq \max_{l \in \mathcal{I}(\tilde{L}_\alpha)} l \max_{r \in [1,l]} \exp\left\{-\frac{(V_l + l\, d_1 - \min(r, k_1)\, q_0(Z) + [l - \max(r, k_1) + 1]_+ q_1(Z))^2}{l \max(\sigma_0^2(Z), \sigma_1^2(Z))}\right\},$$

where $V_l = -\log \Pi_1(l) - l\, d_1$. Note that for $l > L$ for some $L$, $|V_l| < \epsilon$ due to Assumption VII.1. Thus when $r \leq k_1$, the maximum happens at $r = k_1$, with rate upper bounded by

$$r(l) = \left\{\frac{(\epsilon + l\, d_1 - k_1\, q_0(Z) + (l - k_1 + 1)q_1(Z))^2}{l \max(\sigma_0^2(Z), \sigma_1^2(Z))}\right\}.$$

Else, the maximum happens at $r = l$, with rate upper bounded by

$$r(l) = \left\{\frac{(\epsilon + l\, d_1 + q_1(Z))^2}{l \max(\sigma_0^2(Z), \sigma_1^2(Z))}\right\}.$$

In both cases, for any $l \in [k_1 + \tilde{L}_\alpha, \infty]$, $r(l) \to \infty$ as $\alpha \to 0$. Thus we obtain $\mathbb{P}_{k_1,k_2}(\nu_1 > \tilde{\nu}_1, \tilde{\nu}_1 \geq k_1 + \tilde{L}_\alpha) = o(1)$. Since $k_1$ only appears multiplying an exponentially small probability, as both rates go to infinity uniformly over $k_1$, we can apply expectation to both sides, and obtain that $\mathbb{P}_{\lambda_1,\lambda_2}(\nu_1 > \tilde{\nu}_1) = o(1)$, as $\mathbb{E}[k_1] = \lambda_1 < \infty$. ∎